\documentclass[fleqn,twoside]{article}%
\usepackage{amsmath}
\usepackage{amsfonts}
\usepackage{amssymb}
\usepackage{graphicx}%
\def\be{\begin{equation}}

\def\ee{\end{equation}}
\def\bes{\begin{equation}\begin{split}&}
\def\es{\end{split}}
\def\bi{\bibitem}

\begin{document}
\title{Equivalent and inequivalent canonical structures of higher order theories of gravity.}
\author{Ranajit Mandal$^\dag$, Abhik Kumar Sanyal$^\ddag$}

\maketitle
\noindent
\begin{center}
\noindent
$^{\dag}$ Dept. of Physics, University of Kalyani, West Bengal, India - 741235.\\
\noindent
$^{\ddag}$ Dept. of Physics, Jangipur College, Murshidabad,
\noindent
West Bengal, India - 742213. \\

\end{center}
\footnotetext[1] {\noindent
Electronic address:\\
\noindent
$^{\dag}$ranajitmandalphys@gmail.com\\
$^{\ddag}$sanyal\_ ak@yahoo.com}

\abstract{Canonical formulation of higher order theory of gravity can only be accomplished associating additional degrees of freedom, which are extrinsic curvature tensor. Consequently, to match Cauchy data with the boundary data, terms in addition to the three-space metric, must also be fixed at the boundary. While, in all the three, viz. Ostrogradski's, Dirac's and Horowitz' formalisms, extrinsic curvature tensor is kept fixed at the boundary, a modified Horowitz' formalism fixes Ricci scalar, instead. It has been taken as granted that the Hamiltonian structure corresponding to all the formalisms with different end-point data are either the same or are canonically equivalent. In the present study, we show that indeed it is true, but only for a class of higher order theory. However, for more general higher order theories, e.g. dilatonic coupled Gauss-Bonnet gravity in the presence of curvature squared term, the Hamiltonian obtained following modified Horowitz' formalism is found to be different from the others, and is not related under canonical transformation. Further, it has also been demonstrated that although all the formalisms produce viable quantum description, the dynamics is different and not canonically related to modified Horowitz' formalism. Therefore it is not possible to choose the correct formalism which leads to degeneracy in Hamiltonian.}\\

\noindent
keywords: Higher Order theory; Canonical Formulation.\\
\maketitle
\flushbottom
\section{Introduction}

Canonical prescription for higher order theory should  follow the standard canonical quantization scheme, particularly in terms of basic variables - the true degrees of freedom, to be more precise, to produce a hermitian Hamiltonian operator, so that the time evolution of the quantum states is unitary \footnote{This apparently does not make sense in gravity, since the quantum states are devoid of time evolution. However, it make sense as we shall finally produce a Schr\"odinger like equation where an internal variable acts as time parameter}. This may be accomplished by introducing additional degree of freedom. For higher order theory of gravity, in addition to the induced three metric $h_{ij}$, the extrinsic curvature tensor $K_{ij}$, being the basic variable, plays the role. It is thus essential to match Cauchy data with the boundary data. Hence along with $h_{ij}$, additional term is required to keep fixed at the boundary. In most of the techniques existing in the literature, which include  Ostrogrdski's \cite{1O}, Dirac's \cite{2D, 3}, Horowitz' \cite{3H} and Buchbinder-Lyakovich's \cite{4BL, 5BL} techniques, $h_{ij}$ and $K_{ij}$ are kept fixed at the boundary. As a result, the supplementary boundary terms which appear under standard metric variation of the action, disappear. In the process, one also looses the most cherished Gibbons-Hawking-York boundary term \cite{6GH, 7GH, 8Y}, which appears from the linear Einstein-Hilbert sector. Note that for linear gravity, while calculating black-hole entropy using the Euclidean semiclassical approach, the entire contribution comes from the Gibbons-Hawking-York \cite{6GH, 7GH, 8Y} boundary term, and practically, there is no clear physical understanding as to why the concept of black-hole entropy would get lost in strong gravity, i.e. in the presence of higher order terms in the action. Further, it is not possible to retrieve the weak field limit automatically, simply by setting the coupling parameter to the higher order curvature invariant term, $\beta \rightarrow 0$. The reason is that the Gibbons-Hawking-York term \cite{6GH, 7GH, 8Y} does not reappear in the process, unless additional condition is imposed. On the contrary, for higher derivative theory, the derivatives of the metric encode true degrees of freedom, and $\delta(\partial_{\sigma}g_{\mu\nu})$ should not remain arbitrary on the boundary. Instead, $\delta(\partial_{\sigma}g_{\mu\nu})$  must be subject to the constraint that the variation of the four-dimensional Ricci scalar $R$, be held fixed on the boundary. This corresponds to holding the scalar field fixed in the equivalent scalar-tensor theory of $F(R)$ gravity. More precisely, scalar-tensor equivalent forms of higher order ($F(R)$) theory to Jordan's frame of reference is found under re-definition of the field variable, or to Einstein's frame of reference under conformal transformation. To obtain field equations from these scalar-tensor equivalent forms following standard variational principle, it is  required to fix the effective scalar field $\Phi = F'(R)$ in Jordan's frame, or $\tilde\phi = \sqrt{3\over 2\kappa}\ln F'(R)$ in Einstein's frame at the boundary, in addition to the metric. This is equivalent to keep the Ricci scalar $R$ fixed at the boundary, and supplementing the action by a generalized Gibbons-Hawking-York term \cite{6GH, 7GH, 8Y}. It is important to mention that, if $R$ is kept fixed at the boundary, then such a supplementary boundary term reproduces the expected ADM energy \cite{9ADM} upon passing to the Hamiltonian formalism, and the correct expression of entropy of a Schwarzschild black hole may be found in the semiclassical limit \cite{10DH}. A modified version of Horowitz' technique \cite{11A1, 11A11, 11A2, 11A21, 11A22, 11A3, 11A31} follows this later prescription.\\

Despite the fact that the modified Horowitz' prescription \cite{11A1, 11A11, 11A2, 11A21, 11A22, 11A3, 11A31} incorporates all the physical properties required for a viable canonical formulation of higher order theory, it has not been able to draw adequate attention. The reason might be due to the preconceived notion that, all these techniques \cite{1O, 2D, 3H}, either produce the same Hamiltonian, or if different \cite{11A1, 11A11, 11A2, 11A21, 11A22, 11A3, 11A31}, are related under canonical transformation. Indeed it's true, but only for a class of actions containing higher order curvature invariant terms in the presence of minimally coupled scalar-tensor theory of gravity. We demonstrate such equivalence of different phase-space Hamiltonian, under canonical transformation in the following section, in Robertson-Walker minisuperspace model. In section 3, we consider dilatonically coupled Gauss-Bonnet action in the presence of scalar curvature squared term. We then follow all the three formalisms (ostrogradski's \cite{1O}, Dirac's \cite{2D} and Horowitz' \cite{3H}) to show that the phase-space Hamiltonian so obtained are the same, but differ considerably from the one obtained in view of modified Horowitz' prescription \cite{11A1, 11A11, 11A2, 11A21, 11A22, 11A3, 11A31}, since the latter is not related to the former under canonical transformation. It is important to mention that all the Hamiltonian produce correct classical field equations. However, in the quantum domain they might show different behaviour, and one therefore has to pick up the appropriate Hamiltonian which produces a viable quantum description. In subsections 3.3.2 and 3.4.2, we therefore have attempted semiclassical approximation corresponding to the quantized version of the Hamiltonian obtained following the former techniques and the later, respectively. Both the different quantum dynamics produce appropriate although different classical analogues. It is therefore not possible at this stage to choose one of these techniques as an appropriate one, and the issue of boundary fixing is left as a taste. The fact that two different Hamiltonian can describe the same system leads to the pathology of degeneracy in Hamiltonian. We conclude in section 4.

\section{Higher order theory of gravity with equivalent canonical structure}

To demonstrate the fact that despite different choice regarding boundary condition, canonical formulation of the higher order gravitational action produces the same phase-space Hamiltonian, or if different, are canonically equivalent, we consider minimally coupled scalar-tensor theory of gravity in the presence of scalar curvature square ($R^2$) term as \footnote{Addition of $R_{\mu\nu}^2$ term  doesn't make any difference, since in Robertson-Walker metric, the combination $R_{\mu\nu}R^{\mu\nu} - {1\over 3}R^2$ only gives a total derivative term.},
\be\label{AA} A = \int \sqrt{-g} \left[{\alpha}R + {\beta}R^2-\frac{1}{2}\phi_{,\mu}\phi^{,\nu}-V(\phi)\right]d^4x+\alpha\Sigma_{R} +\beta \Sigma_{R^2},\ee
where $\alpha (={1\over 16\pi G})$ and $\beta$ are coupling parameters, $\Sigma_{R}= 2\oint_{\partial\mathcal{V}}K \sqrt hd^3x$ is the Gibbons-Hawking-York supplementary boundary term associated with Einstein-Hilbert action, and $\Sigma_{R^2} = 4\oint_{\partial\mathcal{V}} R K\sqrt h d^3x$ is its modified version corresponding to $R^2$ term, while, $K$ is the trace of the extrinsic curvature tensor $K_{ij}$. Note that, both the counter terms are required if $\delta R = 0$ is fixed at the boundary. However, if $K_{ij}$ instead is fixed at the boundary, as in the case of Ostrogradski's technique \cite{1O}, Dirac constrained analysis \cite{2D}, or Horowitz' formalism \cite{3H}, the counter terms are not required, since both the boundary terms appearing under metric variation vanish. Under metric variation \cite{fld, fld1}, the field equation is obtained as,
\be\label{A1} \alpha \left(R_{\mu\nu}-\frac{1}{2}g_{\mu\nu}R\right)+\beta\left(2RR_{\mu\nu}+2g_{\mu\nu}\Box R-2 \nabla_{\mu} \nabla_{\nu} R-\frac{1}{2} g_{\mu\nu} R^2\right)-T_{\mu\nu}=0\ee
where $T_{\mu\nu}=(\nabla_{\mu}\phi\nabla_{\nu}\phi-\frac{1}{2}g_{\mu\nu}\nabla_{\lambda}\phi\nabla^{\lambda}\phi-g_{\mu\nu}V)$, $\nabla_{\mu}\nabla_{\nu}R=(R_{;\nu})_{;\mu}=\partial_{\mu}(R_{;\nu})-\Gamma^{\beta}_{\mu\nu}R_{;\beta}$, $(R_{;\nu})_{;\mu}=\partial_{\mu}(\partial_{\nu}R)-\Gamma^{\beta}_{\mu\nu}\partial_{\beta}R$, and $\Box R=g^{\mu\nu}\nabla_{\mu}\nabla_{\nu}R=g^{\mu\nu}\partial_{\mu}(\partial_{\nu}R)-g^{\mu\nu}\Gamma^{\beta}_{\mu\nu}\partial_{\beta}R.$ In homogeneous and isotropic Robertson-Walker metric, viz.,
\be\label{RW} ds^2 = -N^2(t)~ dt^2 + a^2(t) \left[\frac{dr^2}{1-kr^2} + r^2 (d\theta^2 + sin^2 \theta d\phi^2)\right],\ee
the Ricci scalar is expressed as,
\be \label{R1}R = \frac{6}{N^2}\left(\frac{\ddot a}{a}+\frac{\dot a^2}{a^2}+N^2\frac{k}{a^2}-\frac{\dot N\dot a}{N a}\right).\ee
Since reduction of higher order theory to its canonical form requires an extra degree of freedom, hence, in addition to the three-space metric $h_{ij}$, the extrinsic curvature tensor $K_{ij}$ is treated as basic variable. We therefore choose the basic variables $h_{ij} = z \delta_{ij} = a^2 \delta_{ij}$, so that $K_{ij} = -{\dot h_{ij}\over 2N} = -{a\dot a\over N} = -\frac{\dot z}{2 N}$. In the above $\delta_{ij}$ is Kronecker delta function. Hence, the Ricci scalar takes the form
\be \label{R2}R = {6\over N^2}\left[{\ddot z\over 2z} + N^2 {k\over z} - {1\over 2}{\dot N\dot z\over N z}\right],\ee
and the action (\ref{AA}) may be expressed as
\be\label{rr}\begin{split}
&A= \int\Big[ {3\alpha\sqrt z}\Big(\frac{\ddot z}{ N}- \frac{\dot N \dot z}{N^2} + 2k N \Big) +\frac{9 \beta}{\sqrt z}\Big(\frac{{\ddot z}^2}{N^3} - \frac{2 \dot N \dot z \ddot z}{N^4} + \frac{{\dot N}^2{\dot z}^2}{N^5} -\frac{4k\dot N \dot z}{N^2}+ \frac{4 k {\ddot z}}{N} + 4 k^2 N \Big)+z^{\frac{3}{2}}\Big(\frac{\dot\phi^2}{2N}-V N\Big)\Big]dt \\&+\alpha \Sigma_R +\beta\Sigma_{R^2}.\end{split}\ee
The $(^0_0)$ component of the field equation in terms of the scale factor $a$ takes the following form
\be\label{A2}\begin{split}& -\frac{6\alpha}{a^2}\Big(\frac{\dot a^2}{N^2}+k\Big)-\frac{36\beta}{a^2N^4}\Big(2\dot a\dddot a-2\frac{\dot a^2\ddot N}{N}-\ddot a^2-4\frac{\dot a\ddot a\dot N}{N}+2\frac{\dot a^2\ddot a}{a}+5\frac{\dot a^2\dot N^2}{N^2}-2\frac{\dot a^3\dot N}{a N}-3\frac{\dot a^4}{a^2}-2\frac{k N^2\dot a^2}{a^2}+\frac{k^2N^4}{a^2}\Big)\\&+\Big(\frac{\dot\phi^2}{2N^2}+V\Big)=0,\end{split}\ee
which when expressed in terms of phase space variables, turns out to be the constrained Hamiltonian of the theory under consideration.

\subsection{Ostrogradski's formalism}

Ostrogradskis formalism gives special treatment to the highest derivatives of the original Lagrangian, so that the initial higher-order regular system be reduced to a first-order system. In Ostrogradski's formalism \cite{1O} if a Lagrangian contains maximum order of $m$-th time derivatives of the generalized coordinate $q_i$, i.e,
\be\label{oos} L = L\left(q_i, \dot q_i, \ddot q_i,....,\stackrel{m}{q_i}\right), \ee
where $\stackrel{m}{q_i} = \left(\frac{d}{dt}\right)^mq_i$, $i=1,2,.....,N $, then one should choose $m$ independent variables $(q_i, \dot q_i, \ddot q_i,....,\stackrel{m-1}{q_i})$ and corresponding $m$ generalized momenta $(p_{i,0}, p_{i,1}, p_{i,2},....,p_{i,m-1})$ according to the following recurrence relation,
\begin{subequations}\begin{align}
& p_{i,m-1} = \frac{\partial L}{\partial q_i^m} \\
& p_{i,n-1} = \frac{\partial L}{\partial q_i^n} - \dot p_{i,n}
\end{align}\end{subequations}
for $n = 1,2,........,m-1$. With these new independent coordinates and their corresponding momenta, one can therefore apply the following Legendre transformation
\be H = \sum _{i=1}^N \sum _{\alpha=0}^{m-1} \dot q_i^\alpha p_{i,\alpha} - L,\ee
to express the Hamiltonian in terms phase space variables. To pursue the technique in the present situation, let us therefore choose an additional variable $x = {\dot z\over N} = -2 K_{ij}$, and express the action (\ref{rr}) in terms of $z$ and $x$ as,
\be\label{O1} A=\int\Big[3\alpha\sqrt z\left(\dot x+2kN\right)+\frac{9\beta}{N\sqrt z}\left(\dot x+2kN\right)^2+ z^{\frac{3}{2}}\Big(\frac{\dot\phi^2}{2N}-VN\Big)\Big] dt. \ee
As already mentioned, the boundary terms don't appear due to the choice $\delta h_{ij} = 0 = \delta K_{ij}$ at the boundary. Now, since the Hessian determinant vanishes, the Lagrangian is singular, and therefore Ostrogradski's prescription cannot be pursued, unless the lapse function $N$ is fixed a-priori. Under the usual choice $N=1$, the action (\ref{O1}) is expressed as,
\be\label{O11} A=\int\Big[3\alpha\sqrt z\left(\dot x+2k\right)+\frac{9\beta}{\sqrt z}\left(\dot x+2k\right)^2+z^{\frac{3}{2}}\Big(\frac{\dot\phi^2} {2}-V\Big)\Big] dt.\ee
The canonical momenta are then,
\be \label{mo} p_x=\frac{\partial L}{\partial \dot x}=3\alpha\sqrt{z}+\frac{18\beta}{\sqrt{z}}\left({\dot x} + 2k  \right);~~~ p_z=\frac{\partial L}{\partial\dot z}-\dot {p_x}=-\frac{18\beta \ddot x}{\sqrt{z}}+\frac{9\beta x}{z^{\frac{3}{2}}}\left({\dot x} + 2k  \right)-\frac{3\alpha x}{2\sqrt{z}};~~~ p_{\phi}=\dot\phi z^{\frac{3}{2}}.\ee
In terms of phase space variables the constrained Hamiltonian is now expressed as
\be \label{HO} {\mathcal H_O}=xp_z+\frac{\sqrt z}{36\beta}p_x^2-2kp_x-\frac{\alpha zp_x}{6\beta} +\frac{\alpha^2z^{\frac{3}{2}}}{4\beta} +\frac{p_{\phi}^2}{2z^{\frac{3}{2}}}+Vz^{\frac{3}{2}}.\ee
One can trivially check that the above Hamiltonian constraint equation leads to the $(^0_0)$ equation of Einstein \eqref{A2}, under the choice $N = 1$.

\subsection{Dirac's constraint analysis}
In the presence of the Lapse function, the Lagrangian corresponding to action \eqref{O1} becomes singular, as mentioned. Therefore, instead of Ostrogradski's technique it is required to follow Dirac's constraint analysis, to cast the action (\ref{rr}) in canonical form. We therefore introduce the constraint $\frac{\dot z}{N}-x=0$ through Lagrange multiplier $\lambda$ in action (\ref{O1}), so that the point Lagrangian now reads,
\be\label{drc}L=
{3\alpha\sqrt z}\big({\dot x} + 2k N \big) +\frac{9 \beta}{N\sqrt z}\big( {\dot x} + 2k N\big)^2+z^{\frac{3}{2}}\Big(\frac{\dot\phi^2}{2N}-VN\Big) +\lambda\Big(\frac{\dot z}{N}-x\Big)\ee
and the corresponding canonical momenta are
\begin{equation}
 p_x=\frac{\partial L}{\partial\dot x}=3{\alpha\sqrt z}+\frac{18\beta}{N\sqrt{z}}\left({\dot x} + 2k N\right);~~
 p_z=\frac{\lambda}{N};~~
 p_{\phi}=\frac{z^{\frac{3}{2}}\dot\phi}{N};~~
 p_N=0;~~
 p_{\lambda}=0.
\end{equation}
The constraint Hamiltonian therefore is,
\be\begin{split}\label{csd} H_c&=\dot x p_x+\dot z p_z+\dot\phi p_{\phi}+\dot N p_N+\dot\lambda p_{\lambda}-L \end{split}\ee
Clearly we require three primary constraints involving Lagrange multiplier or its conjugate viz, $\phi_1= Np_z-{\lambda} \approx 0,\ \,  \ \phi_2=p_{\lambda} \approx 0,\ \,  \text{and},\ \ \phi_3 = p_N \approx 0$, which are second class constraints, as $\{\phi_i,\phi_j\} \ne 0$.
Note that, since the lapse function $N$ is non-dynamical, so the associated constraint vanishes strongly. The first two constraints can now be harmlessly substituted and the modified primary Hamiltonian reads,
\be\begin{split}\label{phd1} H_{p1}&=\frac{N\sqrt z}{36\beta}p_x^2-2 k N p_x-\frac{N\alpha z p_x}{6\beta}+\frac{N\alpha^2z^{\frac{3}{2}}}{4\beta}+\frac{Np_{\phi}^2}{2z^{\frac{3}{2}}}+VNz^{\frac{3}{2}}+\lambda x +u_1\left(Np_z-\lambda\right)+u_2p_{\lambda}.\end{split}\ee
In the above, $u_1\ \text{and} \ u_2$ are Lagrange multipliers, and the Poisson brackets $\{x,p_x\}=\{z,p_z\}=\{\lambda,p_{\lambda}\}=1$, hold. Now constraint should remain preserved in time, which is exhibited in the Poisson brackets $\{\phi_i,H_{pi}\}$ viz,
\be\begin{split}\dot\phi_1 &=\{\phi_1,H_{p1}\}= -N\frac{\partial H_{p1}}{\partial z}-u_2+ \Sigma_{i=1}^2\phi_i\{\phi_1,u_i\},\end{split}\ee
\be \dot\phi_2=\{\phi_2,H_{p1}\}=x-u_1+ \Sigma_{i=1}^2\phi_i\{\phi_2,u_i\}. \ee
Constraints must vanish weakly in the sense of Dirac. As a result,
$\{\phi_1,H_{p1}\}=\dot\phi_1\approx 0$, requires $u_2=-N\frac{\partial H_{p1}}{\partial z}$, and
$\{\phi_2,H_{p1}\}=\dot\phi_2\approx 0,$ requires $u_1 =x$.
On thus imposing these conditions, $H_{p1}$ is then modified by the primary Hamiltonian $H_{p2}$ as
\be\label{rm}\begin{split} H_{p2}&= N x p_z+\frac{N\sqrt z}{36\beta}p_x^2-2kNp_x-\frac{N\alpha z p_x}{6\beta}+\frac{N\alpha^2z^{\frac{3}{2}}}{4\beta}+\frac{Np_{\phi}^2}{2z^{\frac{3}{2}}}+VNz^{\frac{3}{2}}\\&-N\Big(\frac{N}{72\sqrt z\beta}p_x^2-\frac{N\alpha p_x}{6\beta}+\frac{3N\alpha^2\sqrt z}{8\beta} -\frac{3Np_{\phi}^2}{4z^{\frac{5}{2}}}+\frac{3}{2}VN\sqrt{z}\Big)p_{\lambda}.\end{split}\ee
Now, again constraints must vanish weakly in the sense of Dirac, and therefore in view of
$\{\phi_1,H_{p2}\}=\dot\phi_1\approx 0$, one obtains $p_{\lambda}=0$. Thus the Hamiltonian finally takes the form,
\be\label{hd11}  H_D =  N\left[xp_z+\frac{\sqrt z}{36\beta}p_x^2-2kp_x-\frac{\alpha zp_x}{6\beta}+\frac{\alpha^2z^{\frac{3}{2}}}{4\beta} +\frac{p_{\phi}^2}{2z^{\frac{3}{2}}}+Vz^{\frac{3}{2}}\right]=N\mathcal H_D.\ee
Note that the constrained Hamiltonian $\mathcal H_D$ is identical to the one ${\mathcal H_O}$ (\ref{HO}) obtained following Ostrogradski's formalism, setting $N = 1$, a-priori. In view of the Hamilton's equations one obtains,
\be\label{ap}
p_x=3{\alpha\sqrt z}+\frac{18\beta}{N\sqrt{z}}\left({\dot x} + 2k N\right),\;\;\;\dot p_x=-\frac{\partial \mathcal H}{\partial x}=-Np_z,
\ee
and using (\ref{ap}), one finds, \be \begin{split}&\left(\dot z p_z + \dot x p_x +\dot\phi p_{\phi}- N\mathcal{H_D}\right)=\dot x p_x+\dot\phi p_{\phi}-\left[\frac{N\sqrt z}{36\beta}p_x^2-2kNp_x-\frac{N\alpha zp_x}{6\beta}+\frac{N\alpha^2z^{\frac{3}{2}}}{4\beta}+\frac{p_{\phi}^2}{2z^{\frac{3}{2}}}+Vz^{\frac{3}{2}}\right]\\& =3{\alpha\sqrt z\dot x}+\frac{18\beta\dot x}{N\sqrt{z}}\left({\dot x} + 2k N\right)-\left[\frac{9\beta}{N\sqrt z}\left({\dot x} + 2k N\right)^2-6k\alpha N\sqrt z-\frac{36\beta k}{\sqrt z}\left({\dot x} + 2k N\right)\right]+z^{\frac{3}{2}}\left(\frac{\dot\phi^2}{2N}-VN\right)\\&=\Bigg[3\alpha\sqrt z\left(\dot x+2kN\right)+\frac{9\beta}{N\sqrt z}\left(\dot x+2kN\right)^2+z^{\frac{3}{2}}\left(\frac{\dot\phi^2}{2N}-VN\right)\Bigg].\end{split}\ee
Therefore, action (\ref{rr}) can now be expressed in the canonical form in terms of the basic variables as,
\be \label{can}A = \int\left( \dot z p_z + \dot x p_x+\dot\phi p_{\phi} - N\mathcal{H_D} \right)dt~ d^3 x\, \ \
= \int\left(\dot h_{ij} \pi^{ij} + \dot K_{ij}\Pi^{ij} +\dot\phi p_{\phi}- N\mathcal{H_D}\right)dt~ d^3 x,\ee
where, $\pi^{ij}$ and $\Pi^{ij}$ are momenta canonically conjugate to $h_{ij}$ and $K_{ij}$ respectively. Although the Hamiltonian \eqref{hd11} is identical to (\ref{HO}) for $N = 1$, their quantum versions might be appreciably different, since the constraints are second class, as already mentioned. To check, one needs to compute Dirac's bracket (DB) instead of Poisson bracket (PB). Dirac bracket of two functions on phase space, $f$ and $g$, is defined as
\be \label{bdc1} \{f,g\}_{DB}=\{f,g\}_{PB}-\sum_{i,j}\{f,\phi_i\}_{PB}M_{ij}^{-1}\{\phi_j,g\}_{PB},\ee
where, $M_{ij} = \{\phi_i, \phi_j\}_{PB}$, which always has an inverse denoted by $M_{ij}^{-1}$. In the present case, the matrix and its inverse are simply
$M_{ij}=\left( \begin{array}{cc} 0 & -1 \\ 1 & 0
\end{array} \right),$ and $M{ij}^{-1}=\left( \begin{array}{cc} 0 & 1 \\ -1 & 0 \end{array} \right).$
Therefore, the Dirac bracket reduces to the following form
\be \label{bdc2} \{f,g\}_{DB}=\{f,g\}_{PB}+\sum_{i,j}\epsilon_{ij}\{f,\phi_i\}_{PB}\{\phi_j,g\}_{PB},\ee
where $\epsilon_{ij}$ is the Levi-Civita symbol. A straight forward calculation yields
\be \begin{split}& \{z,p_z\}_{DB}=\{z,p_z\}_{PB}+\epsilon_{11}\{z,\phi_1\}_{PB}\{\phi_1,p_z\}_{PB}+\epsilon_{12} \{z,\phi_1\}_{PB}\{\phi_2,p_z\}_{PB}+\epsilon_{21}\{z,\phi_2\}_{PB}\{\phi_1,p_z\}_{PB}\\&+\epsilon_{22}\{z,\phi_2\}_{PB}\{\phi_2,p_z\}_{PB}= \{z,p_z\}_{PB} =1\end{split}.\ee
Likewise, $\{x, p_x\}_{DB} = \{x, p_x\}_{PB} = 1, ~~\{z, p_x\}_{DB} = \{z, p_x\}_{PB} = 0,~~ \{p_z, p_x\}_{DB} = \{p_z, p_x\}_{PB} =0$.
Therefore, the correct implementation of canonical quantization dictates the standard commutation relations, $[\hat z, \hat p_z]=i\hbar = [\hat x, \hat p_x], [\hat p_z, \hat p_x] =0.$ Thus, in the quantum domain too, the Ostrogrdski's and Dirac's formalisms match .

\subsection{Horowitz' formalism}

In contrast to the heuristic speculations made by Boulware \cite{Boul} regarding the structure of momenta and Hamiltonian, Horowitz \cite{Horo} suggested a consistent technique towards canonical formulation of higher order theory. Horowitz \cite{Horo} argued against supplementary boundary terms and insisted on keeping $h_{ij}~\mathrm{and}~K_{ij}$ fixed at the boundary, so that higher order theory is devoid of supplementary boundary terms. However, treating $K_{ij}$ as a variable from the beginning, requires to vary the action with respect to $K_{ij}$ as well, together with $h_{ij}$, since both are treated on the same footing. This restricts classical solutions by and large. Therefore, instead of expressing the action in terms of basic variable $K_{ij}$ a-priori, as in the case of Ostrogradski's and Dirac's technique, Horowitz started with an auxiliary variable $Q_{ij}$, which is found by varying the action with respect to the highest derivative of the field variables present in the action. The Hamiltonian so obtained, was finally expressed in terms of the basic variables $\{h_{ij}, K_{ij}; p^{ij}, \Pi^{ij}\}$, following canonical transformation. The resulting quantized version (the modified Wheeler-deWitt equation) corresponding to the positive definite action obtained in the process, resembles with Schr\"odinger equation, where the internal parameter, viz. the three metric $h_{ij}$ plays the role of time. In this section we apply Horowitz' formalism in connection with the prescribed action (\ref{rr}) in the Robertson-Walker minisuperspace model \eqref{RW}. Introducing the auxiliary variable
\be\label{2.3}
Q ={\partial L\over \partial\ddot z} =\frac{3\alpha\sqrt z}{N}+ \frac{18\beta}{N\sqrt{z}}\Big[\frac{\ddot z}{N^2}-\frac{\dot z\dot N}{N^3}+2k\Big]\ee
judiciously into the action (\ref{rr}), in the following manner
\be\label{3.2}A = \int\Big[ Q \ddot z -\frac{N^3\sqrt{z}}{36\beta}Q^2+2kQN^2+\frac{\alpha N^2Qz}{6\beta}-\frac{\dot N\dot z Q}{N}-\frac{N\alpha^2z^{\frac{3}{2}}}{{4\beta}}+z^{\frac{3}{2}}\Big(\frac{\dot\phi^2}{2N}-VN\Big)\Big]dt\ee
and then removing the boundary terms under integration by parts, one ends up with the following canonical action,
\be\label{3.21}A = \int\Big[- \dot Q \dot z -\frac{N^3\sqrt{z}}{36\beta}Q^2+2kQN^2+\frac{\alpha N^2Qz}{6\beta}-\frac{\dot N\dot z Q}{N}-\frac{N\alpha^2z^{\frac{3}{2}}}{{4\beta}}+z^{\frac{3}{2}}\Big(\frac{\dot\phi^2}{2N}-VN\Big)\Big]dt\ee
Canonical momenta are
\be\label{3.4}
p_{Q} =- \dot z,~~p_{z} =-\dot Q-\frac{\dot N Q}{N},~~p_{\phi}=\frac{z^{\frac{3}{2}}\dot\phi}{N},~~ p_N=-\frac{\dot zQ}{N}\ee
Note that the $Q$ variation equation does not produce any new equation, rather it gives back the definition of the auxiliary variable $Q$ appearing in equation (\ref{2.3}). Therefore, following standard Lengendre transformation, the Hamiltonian,
\be H=\dot Q p_Q+\dot z p_z+\dot\phi p_{\phi}+\dot N p_N-L,\ee
may now be expressed in terms of the phase space variables as,
\be\label{3.51} H_H=- p_Q p_z+\frac{N^3\sqrt{z}}{36\beta}Q^2-2kQN^2-\frac{\alpha QzN^2}{6\beta}+\frac{\alpha^2z^{\frac{3}{2}}N}{4\beta} +\frac{Np_{\phi}^2}{2z^{\frac{3}{2}}} +VNz^{\frac{3}{2}}.\ee
It is now required to express the Hamiltonian in terms of basic variables ($h_{ij},~ K_{ij}$), which in the present case are ($z,~ x = {\dot z\over N} = -2 K_{ij}$). Since, $p_Q =-\dot z=-Nx$ and $Q =\frac{p_x}{N}$ , one therefore is required to replace
$p_Q$ by $-Nx$ and $Q$ by $\frac{p_x}{N}$ in the Hamiltonian (\ref{3.51}). Note that the transformations are canonical. The Hamiltonian therefore reads
\be\label{hh}\begin{split}
H_H &= N\bigg[xp_z +\frac{\sqrt z}{36\beta}p_x^2-2kp_x-\frac{\alpha zp_x}{6\beta}+\frac{\alpha^2z^{\frac{3}{2}}}{4\beta}+\frac{p_{\phi}^2}{2z^{\frac{3}{2}}} +Vz^{\frac{3}{2}} \bigg] = N{\mathcal{H}_H}.
\end{split}\ee
The Hamiltonian so obtained is the same as found in view of Dirac's constrained analysis \eqref{hd11}. The action (\ref{rr}) can therefore be expressed in the canonical form \eqref{can} with respect to the basic variables as well. It is important to note that the canonical momenta ($p_x, p_z, p_{\phi}$) obtained here are the same as those obtained in Ostrogradski's formalism \eqref{mo} under the choice, $N = 1$.

\subsubsection{Quantum counterpart}

It has been established that all the three techniques discussed so far produce the same phase-space Hamiltonian [\eqref{HO} = \eqref{hd11} = \eqref{hh}], at least for the class of higher-order gravitational action \eqref{AA} under consideration. We again mention that addition of $R_{\mu\nu}R^{\mu\nu}$ term makes no difference for the minisuperspace \eqref{RW} under consideration. Let us therefore turn our attention to canonical quantization scheme. Since due to diffeomorphic invariance the Hamiltonian is constrained to vanish, so under canonical quantization one obtains the following modified Wheeler-de-Witt equation,
\be\label{q1}
\frac{i\hbar}{\sqrt z}\frac{\partial \Psi}{\partial z} = -\frac{\hbar^2}{36\beta x}\left(\frac{\partial^2}{\partial x^2} + \frac{n}{x}\frac{\partial}{\partial x}\right)\Psi-\frac{\hbar^2}{2xz^2}\frac{\partial^2\Psi}{\partial\phi^2}+i\hbar\left(\frac{2k}{\sqrt z}+\frac{\alpha \sqrt z}{6\beta }\right)\frac{1}{2}\left(\frac{2}{x}\frac{\partial\Psi}{\partial x}-\frac{\Psi}{x^2}\right)+{z\over x}\left(V(\phi)+\frac{\alpha^2}{4\beta}\right)\Psi\ee
where Weyl symmetric operator ordering has been performed in the 1st. and the 3rd. terms appearing on right hand side, $n$ being the operator ordering index. Under a further change of variable, the above modified Wheeler-de-Witt equation, takes the look of Schr\"odinger equation, viz.,
\be\label{HS} i\hbar \frac{\partial \Psi}{\partial\sigma}=-\frac{\hbar^2}{54\beta }\left(\frac{1}{x}\frac{\partial^2}{\partial x^2} + \frac{n}{x^2}\frac{\partial}{\partial x}\right)\Psi-\frac{\hbar^2}{3x\sigma^{\frac{4}{3}}}\frac{\partial^2\Psi}{\partial\phi^2} +i\hbar\left(\frac{4k}{3\sigma^{\frac{1}{3}}}+\frac{\alpha\sigma^{\frac{1}{3}} }{9\beta }\right)\left(\frac{1}{x}\frac{\partial}{\partial x}-\frac{1}{2 x^2}\right) \Psi+V_e\Psi=\hat{H_e}\Psi\ee
where, the proper volume, $\sigma=z^{\frac{3}{2}}=a^3$ plays the role of internal time parameter. In the above, $\hat H_e$ is the effective Hamiltonian operator and $V_e={2\sigma^{\frac{2}{3}}\over 3x}(V+\frac{\alpha^2 }{4\beta})$ is the effective potential. The hermiticity of the effective Hamiltonian is ensured for $n = -1$, which allows one to write the continuity equation as,
\be   \frac{\partial\rho}{\partial\sigma}+\nabla.\mathbf{J}=0,\ee
where, $\rho=\Psi^{*}\Psi$ and $\mathbf{J} = (J_x, J_{\phi}, 0)$ are the probability density and the current density respectively, with, $J_x = \frac{i\hbar}{54\beta x}(\Psi^{*}_{,x}\Psi-\Psi^{*}\Psi_{,x})-\left(\frac{2k}{3\sigma^{\frac{1}{3}}}+\frac{\alpha\sigma^{\frac{1}{3}} }{18\beta}\right)\frac{\Psi^{*}\Psi}{x}$ and $J_{\phi}=\frac{i\hbar}{2x\sigma^{\frac{4}{3}}}(\Psi^{*}_{,\phi}\Psi-\Psi^{*}\Psi_{,\phi})$.
In the process, probabilistic interpretation becomes straight-forward for higher order theory of gravity under consideration. The reason for taking up the canonical quantization scheme, will be made clear later.

\subsection{Modified Horowitz' formalism}

As already mentioned, Horowitz \cite{Horo} argued against supplementary boundary terms in the higher order theory. His argument ran as follows. Firstly, setting both $\delta h_{ij}|_{\mathcal{\partial V}}=0=\delta K_{ij}|_{\mathcal{\partial V}}$, the solutions of the classical field equation turns out to be the extrema of the action, and so there is no need to add boundary term. Next, without the need of supplementary boundary term, superposition principle holds during the transition from the initial configuration space to the final, following an intermediate one. However, he also pointed out that, the above arguments don't specifically state that boundary terms can't exist. One important consequence of no boundary proposal of Horowitz is that, in the weak field limit, when higher order curvature invariant terms remain subdominant, since the GHY term doesn't exist, one looses the most cherished black hole entropy formula. This was one of the important issues for which Dyer and Hinterbichler \cite{10DH} argued in favour of boundary terms. Their argument run as follows. Firstly, it is well-known that $F(R)$ theory of gravity has scalar tensor equivalence, under redefinition of $F(R)$ by an auxiliary variable to Jordan's frame or through conformal transformation to Einstein's frame. Variation of such canonical Lagrangian clearly requires to fix the scalar at the end point. This is indeed equivalent to fix the Ricci scalar $R$ at the boundary, which requires supplementary boundary terms. One can't therefore fix $K_{ij}$ at the end points, since boundary data would then exceed Cauchy data. Next, Dyer and Hinterbichler \cite{10DH} have shown that the boundary terms reproduces the expected ADM energy. They further calculated the entropy of a Schwarzschild black hole and found that it is one-quarter of the area of the horizon in units of the effective Planck's length, which agrees with the Wald entropy formula. This result clearly indicates that higher curvature terms make no correction to the entropy since only GHY term is required. However, such cherished GHY term vanishes if $K_{ij}$ is fixed at the boundary, and it is not possible to recover it under weak field limit, as already mentioned. In this context, we have modified Horowitz' technique earlier, keeping supplementary boundary terms and setting $\delta g_{\mu\nu} = 0 =\delta R$ at the boundary. Since this technique already exists in the literature, here we briefly demonstrate the procedure.\\

We start with the action \eqref{rr}, fix $\delta g_{\mu\nu} = 0 =\delta R$ at the boundary, remembering that the supplementary boundary terms don't vanish trivially any more. It has been noticed in our earlier works \cite{11A1,11A11,11A2,11A21,11A22,11A3,11A31} that the supplementary boundary terms can be taken care of appropriately, only if one integrates out all the existing total derivative terms appearing in the action, prior to the introduction of auxiliary variable. Therefore, we split the boundary term associated with $R^2$ term in two parts, viz.,
\be\label{sigma}\Sigma_{R^2} = \Sigma_{R^2_1} + \Sigma_{R^2_2} = 4\oint_{\partial\mathcal{V}} R K\sqrt h d^3x= 4\oint_{\partial\mathcal{V}} {^3R} K \sqrt h d^3x + 4\oint_{\partial\mathcal{V}} (R - {^3R}) K\sqrt h d^3x,\ee
where ${^3R}$ is the Ricci scalar in 3-space. Under integration by parts, the two of the total derivative terms get cancelled with the supplementary boundary terms $\Sigma_R$ and $\Sigma_{R^2_1}$ and the action (\ref{rr}) takes the following form,

\be \label{rrr} A = \int\Big[\alpha\Big( - \frac{3 {\dot z}^2}{2 N \sqrt z} + 6 k N \sqrt z \Big) +\frac{9 \beta}{\sqrt z}\Big(\frac{{\ddot z}^2}{N^3} - \frac{2 \dot N \dot z \ddot z}{N^4} + \frac{{\dot N}^2{\dot z}^2}{N^5} +\frac{2k {\dot z}^2}{N z}+ 4 k^2 N \Big)+z^{\frac{3}{2}}\Big(\frac{\dot\phi^2} {2N}-VN\Big)\Big]dt  +\beta\Sigma_{R^2_2}.\ee
Now introducing an auxiliary variable $\mathcal Q$, following Horowitz' prescription
\be\label{auv} \mathcal Q=\frac{18\beta}{N^3\sqrt{z}}\Big(\ddot z-{\dot z\dot N\over N}\Big)\ee
judiciously into the action (\ref{rrr}), as
\be\label{la2}A = \int\left[{\mathcal Q} \ddot z-\frac{N^3\sqrt{z}{\mathcal Q}^2}{36\beta}-\frac{\dot z\dot N {\mathcal Q}}{N}-\frac{3\alpha\dot{z}^2}{2N\sqrt z}+ \frac{18\beta k\dot z^2}{Nz^{\frac{3}{2}}} + \frac{36{\beta}Nk^{2}}{\sqrt{z}}+6\alpha kN\sqrt{z}+z^{\frac{3}{2}}\Big(\frac{\dot\phi^2} {2N}-VN\Big)\right]dt+\beta\Sigma_{R^2_2}\ee
one can integrate the action by parts again, to take care of the rest of the boundary terms. The canonical action therefore reads
\be\label{la3}A = \int\left[-\dot{\mathcal Q}\dot z-\frac{N^3\sqrt{z}{\mathcal Q}^2}{36\beta}-\frac{\dot z\dot N {\mathcal Q}}{N}-\frac{3\alpha\dot{z}^2}{2N\sqrt z}+ \frac{18\beta k\dot z^2}{Nz^{\frac{3}{2}}} + \frac{36{\beta}Nk^{2}}{\sqrt{z}}+6\alpha kN\sqrt{z}+z^{\frac{3}{2}}\Big(\frac{\dot\phi^2} {2N}-VN\Big)\right]dt.\ee
The canonical momenta are
\be\label{mo} p_{\mathcal Q}=-\dot z,\;\ p_z=-\dot{\mathcal Q}-\frac{\dot N {\mathcal Q}}{N}-{3\alpha\dot{z}\over N\sqrt z}+\frac{36\beta k\dot z}{Nz^{\frac{3}{2}}},\;\ p_{\phi}=z^{\frac{3}{2}}\frac{\dot\phi^2}{N},\;\ p_N=-{\dot z {\mathcal Q}\over N}.\ee
Under Legendre transformation, the Hamiltonian, $H=\dot{\mathcal Q} p_{\mathcal Q}+\dot z p_z+\dot\phi p_{\phi}+\dot N p_N-L$, may now be expressed in terms of the phase space variables as,
\be\label{mdh} H_L=-p_{\mathcal Q} p_z+\frac{N^3\sqrt{z}}{36\beta}{\mathcal Q}^2+\frac{3\alpha p_{\mathcal Q}^2}{2N\sqrt z}-\frac{18\beta k p_{\mathcal Q}^2}{Nz^{\frac{3}{2}}}-\frac{36{\beta}Nk^{2}}{\sqrt{z}}-6\alpha kN\sqrt{z}+\frac{N p_{\phi}^2}{2z^{\frac{3}{2}}}+VNz^{\frac{3}{2}}\ee
Finally, under the canonical transformation, ${\mathcal Q}=\frac{p_x}{N},~ \mathrm{and},~p_{\mathcal Q}=-\dot z=-Nx$, the above Hamiltonian may be expressed in terms of basic variables as,

\be\label{hhl2} H_L=N\Big[x p_z+\frac{\sqrt{z}}{36\beta}p_x^2+\frac{3\alpha {x}^2}{2\sqrt z}-\frac{18\beta kx^2}{z^{\frac{3}{2}}} -\frac{36{\beta}k^{2}}{\sqrt{z}}-6\alpha k \sqrt{z}+\frac{p_{\phi}^2}{2z^{\frac{3}{2}}}+Vz^{\frac{3}{2}}\Big] = N{\mathcal{H}} = 0.\ee
The action (\ref{rr}) may now be expressed in the canonical form with respect to the basic variables as,
\be\ A = \int\left(\dot z p_z + \dot x p_x +\dot\phi p_{\phi}- N\mathcal{H}\right)dt~ d^3 x = \int\left(\dot h_{ij} \pi^{ij}  + \dot K_{ij}\Pi^{ij} ++\dot\phi p_{\phi}- N\mathcal{H}\right)dt~ d^3 x\ee
where, the second equality is an outcome of the fact that $\dot x = -2 {\dot K_{ij}}$, and, $p_x = -{\frac{1}{2}}\Pi_{ij}$.
Although, the above Hamiltonian (\ref{hhl2}) looks different, under the following canonical transformations
\be\label{tr} p_z= P_Z-18\beta\frac{kX}{Z^{\frac{3}{2}}}+\frac{3\alpha X}{2\sqrt{Z}};~~ z=Z,\;\;\;\;\;p_x=P_X+\frac{36\beta k}{\sqrt{Z}}-3\alpha\sqrt{Z};~~~ x=X,\;\;\; \mathrm{and}\;\;\; p_{\phi}=P_{\Phi};~~~~\phi=\Phi\ee
it reduces to one and the same form as obtained earlier \eqref{HO}, \eqref{hd11}, \eqref{hh}. This proves equivalence of different techniques towards canonical formulation of higher order theory. It is important to mention that all the Hamiltonian obtained following different routes towards canonical formulation, when expressed in terms of the scale factor $a$, and divided by $\sqrt{-g} = Na^3 $, the $(^0_0)$ equation of Einstein (\ref{A2}) is retrieved.

\subsubsection{Quantum counterpart}
It is interesting to note that the choice of the boundary term doesn't tell upon the Hamiltonian structure of the theory. However, there is a shuttle advantage to follow the modified Horowitz' scheme, that we explore here. Under standard canonical quantization scheme, the Hamiltonian (\ref{hhl2}) produces the following modified Wheeler-De-Witt equation,

\be\label{q2}
\frac{i\hbar}{\sqrt z}\frac{\partial \Psi}{\partial z} = -\frac{\hbar^2}{36\beta x}\left(\frac{\partial^2}{\partial x^2} + \frac{n}{x}\frac{\partial}{\partial x}\right)\Psi-\frac{\hbar^2}{2xz^2}\frac{\partial^2\Psi}{\partial\phi^2}-V_e\Psi,\ee
In the above, $V_e$ is the effective potential term given by,
 \be V_e=\frac{Vz}{x}+\frac{3\alpha x}{2z}-\frac{18\beta k x}{z^2}-\frac{36\beta k^2}{xz}-\frac{6\alpha k}{x}.\ee
Weyl symmetric ordering has been performed in the 1st. term appearing on the right hand side of (\ref{q2}), $n$ being the operator ordering index. Now, again under a further change of variable, the above modified Wheeler-de-Witt equation, takes the look of Schr\"{o}dinger
equation, viz.,
\be\label{HS1} i\hbar \frac{\partial \Psi}{\partial\sigma}=-\frac{\hbar^2}{54\beta }\left(\frac{1}{x}\frac{\partial^2}{\partial x^2} + \frac{n}{x^2}\frac{\partial}{\partial x}\right)\Psi-\frac{\hbar^2}{3 x\sigma^{\frac{4}{3}}}\frac{\partial^2\Psi}{\partial\phi^2}-\frac{2V_e}{3}\Psi = \hat H_e\Psi,\ee
where, $\sigma=z^{\frac{3}{2}}=a^3$ plays the role of internal time parameter, and $\hat H_e$ is the effective Hamiltonian operator. The hermiticity of the effective Hamiltonian allows one to write the continuity equation for $n = -1$, as,
\be   \frac{\partial\rho}{\partial\sigma}+\nabla.\mathbf{J}=0,\ee
where, $\rho=\Psi^{*}\Psi$ and $\mathbf{J} = (J_x, J_{\phi}, 0)$ are the probability density and the current density respectively, with, $J_x = \frac{i\hbar}{54\beta x}(\Psi^*_{,x}\Psi-\Psi^*\Psi_{,x})$ and $ J_{\phi}=\frac{i\hbar}{2x\sigma^{\frac{4}{3}}}(\Psi^*_{,\phi}\Psi-\Psi^*\Psi_{,\phi})$.\\

The advantage to follow the present technique is apparent as one can now extremize the effective potential at the energy scale where potential energy dominates over the kinetic energy. This leads to the solution of the scale factor for curvature parameter $k=0$, in the form $a\sim e^{\pm \sqrt{\frac{V}{6\alpha}}t}$, which assures inflation at the early stage of cosmic evolution, and is the primary feature of curvature squared gravity. However, in the following section we shall study another class of higher order theory, which produces different phase-space Hamiltonian, not related under canonical transformation.

\section{Higher order theory of gravity with inequivalent canonical structure}

It's true that Hamiltonian for curvature squared gravity as appeared in action \eqref{AA}, obtained following Ostrogrdski's, Dirac's or Horowitz' techniques is related to the one found following the so-called modified Horowitz' technique under canonical transformation. It therefore appears that boundary term doesn't tell upon the Hamiltonian structure. However, this is not true in general, and the difference becomes appearnt for a more general gravitational action. Here, we take up Gauss-Bonnet-Dilatonic coupled higher order gravitational action to explore the difference.\\

Gauss-Bonnet-dilatonic coupled term arises naturally as the leading order of the $\alpha'$ expansion of heterotic superstring theory, where, $\alpha'$ is the inverse string tension \cite{22, 23, 24, 25, 26, 27}. However, canonical formulation of corresponding action suffers from the issue of branching, due to the presence of quartic power of velocity in the Lagrangian, which makes the theory intrinsically nonlinear. All the attempts made so far to resolve the issue, produce different phase-space Hamiltonian, which are not related through canonical transformation. Recently, it has been observed that the addition of curvature squared term allows to bypass the issue of branching both in Lanczos-Lovelock gravity \cite{LL} and Gauss-Bonnet-dilatonic coupled gravity \cite{GBD} successfully. It is important to mention that Gauss-Bonnet-dilatonic coupled gravity has been found to play the role of dark energy \cite{ifgb, lt} allowing crossing to phantom divide line \cite{pdl} at the late stage of cosmological evolution, after transition from a long Friedmann-like matter dominated era. If Gauss-Bonnet-dilatonic coupled term plays its role at the late stage of cosmic evolution, then, Inflation at the very early stage must be the outcome of a different curvature invariant term. What can then be better than the scalar curvature squared term, which invokes Inflation without phase transition \cite{staro}? It is therefore important to try for canonical formulation of the action containing dilatonic coupled Gauss-Bonnet term in the presence of scalar curvature invariant term, taken in the following form,

\be\label{Aa1} A =\int\sqrt{-g}\;d^4x\left[\alpha{R}+\beta R^2+\Lambda(\phi)\mathcal{G}-\frac{1}{2}\phi_{,\mu}\phi^{,\nu}-V(\phi)\right] + \alpha\Sigma_R + \beta\Sigma_{R^2} + \Lambda(\phi)\Sigma_\mathcal{G}.\ee
In the above, $\Sigma_\mathcal{G} = 4\oint_{\partial\mathcal{V}} \left( 2G_{ij}K^{ij} + \frac{\mathcal{K}}{3}\right)\sqrt hd^3x$ is the supplementary boundary term required for Gauss-Bonnet-dilatonic coupled sector, $\Lambda(\phi)$ is the coupling parameter and $V(\phi)$ is the dilatonic potential. The symbol $\mathcal{K}$ stands for $\mathcal{K}=K^3 - 3K K^{ij}K_{ij} + 2K^{ij}K_{ik}K^k_j$. Under variation \cite{fld}, the Field equation is found as,
\be\begin{split}\label{A1} \alpha G_{\mu\nu}-T_{\mu\nu}&+\beta\left(2RR_{\mu\nu}+2g_{\mu\nu}\Box R-2 \nabla_{\mu} \nabla_{\nu} R-\frac{1}{2} g_{\mu\nu} R^2\right)+\Lambda H_{\mu\nu}\\&+4\left(\Lambda'^2\nabla^{\rho}\phi\nabla^{\sigma}\phi-
\Lambda'\nabla^{\rho}\phi\nabla^{\sigma}\phi\right)P_{\mu\rho\nu\sigma}=0\end{split}\ee

\be \mathrm{where},\begin{split}& H_{\mu\nu}=2\left(RR_{\mu\nu}-2R_{\mu\rho}R^{\rho}_{\nu}-2R_{\mu\rho\nu\sigma}R^{\rho\sigma}
+R_{\mu\rho\sigma\lambda}R^{\sigma\rho\lambda}_{\nu}\right)-\frac{1}{2} g_{\mu\nu}\mathcal G\\&
P_{\mu\nu\rho\sigma}=R_{\mu\nu\rho\sigma}+2g_{\mu[\sigma} {R_{\rho]}}_\mu+2g_{\nu[\rho} {R_{\sigma]}}_\mu+ R g_{\mu[\rho} {R_{\sigma]}}\nu\end{split}\ee
In the homogeneous and isotropic Robertson-Walker metric \eqref{RW}, the expression for the Ricci scalar has already been presented in \eqref{R1} in terms of the scale factor and also in  \eqref{R2} in terms of the basic variable. The expression for the Gauss-Bonnet term is,
\be \mathcal{G} = \frac{24}{N^3 a^3}\Big(N\ddot a - \dot N \dot a\Big)\Big(\frac{\dot a^2}{N^2} + k\Big) = {12\over N^2}\Big({\ddot z\over z}-{1\over 2}{\dot z^2\over z^2}-{\dot N\over N}{\dot z\over z}\Big)\Big({1\over 4 N^2}{\dot z^2\over z^2} + {k\over z}\Big).\ee
The $\phi$ variation equation and the $(^0_0)$ component of the Einstein's field equation in terms of the scale factor are,

\be\label{phi} - 24\left({\dot a^2\ddot a\over N^3} + k{\ddot a\over N} - {\dot a^3\dot N\over N^4} - {k\dot a\dot N\over N^2}\right)\Lambda'+{a^3\over N}\left(\ddot \phi + 3 {\dot a\over a}\dot\phi+N^2 V' -{\dot N\over N}\dot\phi\right) = 0.\ee
\be\begin{split}\label{A22}& -\frac{6\alpha}{a^2}\Big(\frac{\dot a^2}{N^2}+k\Big)-\frac{36\beta}{a^2N^4}\Bigg(2\dot a\dddot a-2\dot a^2\frac{\ddot N}{N}-\ddot a^2-4\dot a\ddot a\frac{\dot N}{N}+2\dot a^2\frac{\ddot a}{a}+5\dot a^2\frac{\dot N^2}{N^2}
\\&-2\frac{\dot a^3\dot N}{a N}-3\frac{\dot a^4}{a^2}-2kN^2\frac{\dot a^2}{a^2}+\frac{k^2N^4}{a^2}\Bigg)+\Big(\frac{\dot\phi^2}{2N^2}+V\Big)-\frac{24\Lambda'\dot a\dot\phi}{N^2a^3}\Big(\frac{\dot a^2}{N^2}+k\Big)=0.\end{split}\ee
The action (\ref{Aa1}) in terms of the basic variable $z$ takes the form,
\be\begin{split}\label{A12}
A &= \int\Bigg[{3\alpha\sqrt z}\Big(\frac{\ddot z}{ N}- \frac{\dot N \dot z}{N^2} + 2k N \Big)+\frac{9\beta}{\sqrt z}\Big(\frac{{\ddot z}^2}{N^3} - \frac{2 \dot N \dot z \ddot z}{N^4} + \frac{{\dot N}^2{\dot z}^2}{N^5} -\frac{4k\dot N \dot z}{N^2}+ \frac{4 k {\ddot z}}{N} + 4 k^2 N\Big) \\& + \frac {3 \Lambda (\phi)}{N \sqrt z}\Big(\frac{{\dot z}^2 \ddot z}{N^2 z} + 4 k \ddot z - \frac{{\dot z}^4}{ 2 N^2 z^2} - \frac{\dot N {\dot z}^3}{N^3 z} - \frac{2 k {\dot z}^2}{z}- \frac{4 k \dot N \dot z}{N}\Big)+z^{\frac{3}{2}}\Big(\frac{1}{2N}\dot\phi^2-VN\Big)\Bigg]dt\\&+\alpha\Sigma_R+ \beta\Sigma_{R^2} + \Lambda(\phi)\Sigma_\mathcal{G},\;\;\;\;\;\;\;\;\;
\mathrm{where,}\;\;  \Sigma_\mathcal{G} = -\frac {\dot z}{N \sqrt z}\Big(\frac{{\dot z}^2}{N^2 z} + 12 k\Big).\end{split}\ee

\subsection{Ostrogradski's formalism}
In this subsection, our aim is to find the phase-space structure of the Hamiltonian, corresponding to the action \eqref{Aa1} in Robertson-Walker minisuperspace \eqref{RW}. We therefore start with the above form of the action \ref{A12}. As already mentioned, in all the three techniques (Ostrogradski's, Dirac's and Horowitz') both $h_{ij}$ and $K_{ij}$ are fixed at the boundary and so the supplementary boundary terms vanish trivially. To proceed with canonical formalism following Ostrogradski's or Dirac's technique, it is required to express the action in terms of both the basic variables $h_{ij}, K_{ij}$ a-priori. As before, let us therefore consider $x={\dot z\over N} = -2K_{ij}$ so that the action (\ref{A12}) may be expressed as,
\be\begin{split}\label{o1} A=\int\Bigg[3\alpha\sqrt z\left(\dot x+2kN\right)+\frac{9\beta}{N\sqrt z}\left(\dot x+2kN\right)^2&+\frac{3\Lambda(\phi)}{\sqrt z}\Big[\left(\frac{x^2}{z}+4k\right)\dot x-\frac{x^4N}{2z^2}-\frac{2kNx^2}{z}\Big]\\&+z^{\frac{3}{2}}\left(\frac{1}{2N}\dot\phi^2-V N\right)\Bigg] dt,\end{split}\ee
It has already been noticed that in the presence of the lapse function ($N$), the Lagrangian becomes singular and so Ostrogradski's technique is not admissible. However, fixing the gauge $N=1$, the action (\ref{o1}) becomes non-singular and canonical formulation is possible as before. Canonical momenta are found as,
\be p_x=\frac{\partial L}{\partial \dot x}=3\alpha\sqrt{z}+\frac{18\beta}{\sqrt{z}}\left({\dot x} + 2k  \right) +\frac{3\Lambda(\phi)}{\sqrt z}\left(\frac{x^2}{z}+4k\right)~~;~~ p_z=\frac{\partial L}{\partial\dot z}-\dot {p_x} ~~~~\text{and}~~ p_{\phi}=\dot\phi z^{\frac{3}{2}}.\ee
The phase space structure of the Hamiltonian may then be obtained in a straight forward manner as,
\be\begin{split} \label{ho1} &{\mathcal H_O}=x p_z+\frac{\sqrt{z}p_x^2}{36\beta}-\left(\frac{\alpha z}{6\beta}+\frac{\Lambda(\phi)x^2}{6\beta z}+\frac{2k\Lambda(\phi)}{3\beta}+2k\right)p_x+\frac{p_{\phi}^2}{2z^{\frac{3}{2}}} +\left(\frac{\Lambda^2(\phi)}{4\beta z^{\frac{5}{2}}}+\frac{3\Lambda(\phi)}{2z^{\frac{5}{2}}}\right)x^4\\&+\frac{\alpha^2 z^{\frac{3}{2}}}{4\beta}+\left(\frac{\alpha \Lambda(\phi)}{2\beta \sqrt{z}}+\frac{12k\Lambda(\phi)}{z^{\frac{3}{2}}} +\frac{2k\Lambda^2(\phi)}{\beta z^{\frac{3}{2}}}\right)x^2+\frac{2\alpha k\Lambda(\phi)\sqrt{z}}{\beta}+\frac{24k^2\Lambda(\phi)}{\sqrt z}+\frac{4k^2\Lambda^2(\phi)}{\beta \sqrt{z}}+Vz^{\frac{3}{2}}.\end{split}\ee
Following and involved and tedious calculation, it is possible to confirm that, the above Hamiltonian \eqref{ho1} is indeed the $(^0_0)$ equation of Einstein \eqref{A22} for $N = 1$, when expressed in terms of configuration space variables.

\subsection{Dirac's constraint analysis}
As mentioned, in the presence of lapse function, Lagrangian becomes singular, and so it is required to follow Dirac's constraint analysis. Let us therefore introduce the constraint $({\dot z\over N}-x)=0$ through Lagrange multiplier $\lambda$ in the action (\ref{o1}) as before, so that the point Lagrangian takes the form
\be\begin{split}\label{drc1}L&=3\alpha\sqrt z\left(\dot x+2kN\right)+\frac{9\beta}{N\sqrt z}\left(\dot x+2kN\right)^2+\frac{3\Lambda(\phi)}{\sqrt z}\left(\frac{x^2}{z}+4k\right)\dot x-\frac{3\Lambda(\phi)}{\sqrt z} \left(\frac{x^4N}{2z^2}+\frac{2kNx^2}{z}\right)\\&+z^{\frac{3}{2}} \left(\frac{1}{2N}\dot\phi^2-VN\right)+\lambda\left({\dot z\over N}-x\right)\end{split}\ee
Canonical momenta are
\be\label{dm1} p_x=3\alpha\sqrt{z}+\frac{18\beta}{N\sqrt z}(\dot x+2kN)+ \frac {3\Lambda(\phi)}{\sqrt z}\Big(\frac{x^2}{z}+4k\Big),\;\ p_z={\lambda\over N},\;\ p_{\phi}=\frac{\dot\phi}{N}z^{\frac{3}{2}}\;\ p_N=0\;\ p_{\lambda}=0\ee
The constraint Hamiltonian is
\be\label{csd1} H_{c}=\dot xp_{ x}+\dot zp_z+\dot \phi p_\phi+\dot Np_N+\dot\lambda p_{\lambda}-L.\ee
Since the lapse function $N$ is not a dynamical variable, so $p_N=0.$ Clearly we require the following primary constraints involving Lagrange multiplier or its conjugate viz,
\be \phi_1= N p_z-\lambda, \phi_2 = p_{\lambda},\ee
which are second class, since $\{\phi_1, \phi_2\} \ne 0$. The primary Hamiltonian now takes the form,
\be\begin{split}\label{phd1} H_{p1}&= \frac{N\sqrt{z}}{36\beta}p_x^2+ \frac{N\alpha^2 z^{\frac{3}{2}}}{4\beta}+\frac{\Lambda(\phi)^2N}{4\beta\sqrt z} \Big(\frac{x^2}{z}+4k\Big)^2 -\frac{N\alpha zp_x}{6\beta}-2kNp_x-\frac{\Lambda(\phi)N}{6\beta} \Big(\frac{x^2}{z}+4k\Big)p_x\\&+\frac{\alpha\Lambda(\phi)N\sqrt z}{2\beta} \Big(\frac{x^2}{z}+4k\Big) +\frac{6\Lambda(\phi)k N}{\sqrt z} \Big(\frac{x^2}{z}+4k\Big)+\frac{6\Lambda(\phi)kNx^2}{z^{\frac{3}{2}}}+\frac{3\Lambda(\phi)Nx^4}{2z^{\frac{5}{2}}}\\&+\frac{Np_{\phi}^2}{2z^{\frac{3}{2}}} +VNz^{\frac{3}{2}}+\lambda x+u_1\big(Np_z-{\lambda}\big)+u_2p_{\lambda}.\end{split}\ee
In the above expression, $u_1, u_2$  are the Lagrange multipliers,  and the Poisson bracket $\{x,p_x\}=\{z,p_z\}=\{\lambda,p_{\lambda}\}=1$, hold. Now constraints should remain preserved in time, which are exhibited in the Poisson brackets $\{\phi_i,H_{pi}\}$ viz,
\be\begin{split}\dot\phi_1 &=\{\phi_1,H_{p1}\}=-N\frac{\partial H_{p1}}{\partial z}-u_2+ \Sigma_{i=1}^2\phi_i\{\phi_1,u_i\} \end{split}.\ee
\be\dot\phi_2=-x+u_1+\Sigma_{i=1}^2\phi_i\{\phi_2,u_i\} \ee
Consequently, all the Poisson bracket relations vanish weakly if we set,
\be\begin{split}u_2&= -N\frac{\partial H_{p1}}{\partial z}\end{split}\;\;\; \mathrm{ and}\;\;\;u_1=x,\ee
The Hamiltonian finally takes the following form,
\be\begin{split}\label{Hl2}& H_D=N\Big[x p_z+\frac{\sqrt{z}p_x^2}{36\beta}-\left(\frac{\alpha z }{6\beta}+\frac{\Lambda(\phi)x^2 }{6\beta z}+\frac{2k\Lambda(\phi)}{3\beta}+2k\right)p_x+\frac{p_{\phi}^2}{2z^{\frac{3}{2}}}+\left(\frac{\Lambda^2(\phi)}{4\beta z^{\frac{5}{2}}} +\frac{3\Lambda(\phi)}{2z^{\frac{5}{2}}}\right) x^4+\frac{\alpha^2 z^{\frac{3}{2}}}{4\beta}\\&+\left(\frac{\alpha \Lambda(\phi)}{2\beta \sqrt{z}}+\frac{12k\Lambda(\phi)}{z^{\frac{3}{2}}} +\frac{2k\Lambda^2(\phi)}{\beta z^{\frac{3}{2}}}\right)x^2+\frac{2\alpha k\Lambda(\phi)\sqrt{z}}{\beta}+\frac{24 k^2\Lambda(\phi)}{\sqrt z}+\frac{4k^2\Lambda^2(\phi) }{\beta \sqrt{z}} +Vz^{\frac{3}{2}}\Big] = N{\mathcal{H}_D}.\end{split}\ee
The action (\ref{A12}) can now be expressed in the canonical form with respect to the basic variables as,
\be\begin{split} A &= \int\left(\dot z p_z + \dot x p_x - N\mathcal{H}_D\right)dt~ d^3 x
= \int\left(\dot h_{ij} \pi^{ij} + \dot K_{ij}\Pi^{ij} - N\mathcal{H}_D\right)dt~ d^3 x,\end{split}\ee
where, $\pi^{ij}$ and $\Pi^{ij}$ are momenta canonically conjugate to $h_{ij}$ and $K_{ij}$ respectively. Note that ${\mathcal{H}_D}$ obtained in \eqref{A12} is that same as the one \eqref{ho1} obtained following Ostrgradski's technique. As in the case studied in section (2.2), one can easily check that, the Dirac and the Poisson brackets between the phase space variables are identical here too, and so, standard commutation relations hold. Thus the resulting quantum dynamics following Ostrogradski's technique and Dirac's formalism match.

\subsection{Horowitz' formalism}

Instead of the true degree of freedom viz., $K_{ij}$, Horowitz' technique initiates with an auxiliary variable, found by varying the action with respect to the highest derivative appearing in it. Thus one has to start with action (\ref{A12}), without the supplementary boundary terms, since they vanish under the boundary condition $\delta h_{ij} = 0 = \delta K_{ij}$. Introducing the auxiliary variable
\be\label{Q} Q={\partial L\over \partial\ddot z} =\frac{3\alpha\sqrt z}{N}+ \frac{18\beta}{N\sqrt{z}}\Big(\frac{\ddot z}{N^2}-\frac{\dot z\dot N}{N^3}+2k\Big)+\frac{3\Lambda(\phi)}{N\sqrt z}\Big(\frac{\dot z^2}{N^2 z} + 4 k\Big)\ee
straight into the action (\ref{A12}) as,
\be\begin{split}\label{7}
A &= \int\Bigg[Q\ddot z-\frac{N^3\sqrt{z}Q^2}{36\beta}-\frac{N\alpha^2 z^{\frac{3}{2}}}{4\beta}-\frac{\Lambda(\phi)^2 \dot z^4}{4N^3\beta z^{\frac{5}{2}}} -\frac{4N\Lambda(\phi)^2 k^2}{\beta \sqrt{z}}+\frac{N^2\alpha Qz}{6\beta}+2kQN^2\\&+\frac{\Lambda(\phi)\dot z^2 Q}{6\beta z}+\frac{2kN^2\Lambda(\phi)Q}{3\beta }-\frac{\alpha \Lambda(\phi)\dot z^2 }{2\beta N\sqrt{z}}-\frac{Q\dot N\dot z}{N}-\frac{2k\alpha N\Lambda(\phi)\sqrt {z}}{\beta}\\&-\frac{12k\Lambda(\phi)\dot z^2}{Nz^{\frac{3}{2}}}-\frac{2k\Lambda(\phi)^2\dot z^2}{N\beta z^{\frac{3}{2}}}-\frac{24k^2N\Lambda(\phi)}{\sqrt z}-\frac{3\Lambda(\phi)\dot z^4}{2N^3z^{\frac{5}{2}}}+\frac{z^{\frac{3}{2}}\dot\phi^2}{2N}-VNz^{\frac{3}{2}}\Bigg]dt,\end{split} \ee
and integrating the action \eqref{7} by parts, one ends up with the following canonical action,
\be\begin{split}\label{71}
A &= \int\Bigg[-\dot Q\dot z-\frac{N^3\sqrt{z}Q^2}{36\beta}-\frac{N\alpha^2 z^{\frac{3}{2}}}{4\beta}-\frac{\Lambda(\phi)^2 \dot z^4}{4N^3\beta z^{\frac{5}{2}}} -\frac{4N\Lambda(\phi)^2 k^2}{\beta \sqrt{z}}+\frac{N^2\alpha Qz}{6\beta}+2kQN^2\\&+\frac{\Lambda(\phi)\dot z^2 Q}{6\beta z}+\frac{2kN^2\Lambda(\phi)Q}{3\beta}-\frac{\alpha \Lambda(\phi)\dot z^2 }{2\beta N\sqrt{z}}-\frac{Q\dot N\dot z}{N}-\frac{2k\alpha N\Lambda(\phi)\sqrt {z}}{\beta}\\&-\frac{12k\Lambda(\phi)\dot z^2}{Nz^{\frac{3}{2}}}-\frac{2k\Lambda(\phi)^2\dot z^2}{N\beta z^{\frac{3}{2}}}-\frac{24k^2N\Lambda(\phi)}{\sqrt z}-\frac{3\Lambda(\phi)\dot z^4}{2N^3z^{\frac{5}{2}}}+\frac{z^{\frac{3}{2}}\dot\phi^2}{2N}-VNz^{\frac{3}{2}} \Bigg]dt.\end{split} \ee
Canonical momenta are

\[p_z =-\dot Q + \Big(\frac{\Lambda(\phi)\dot z}{3\beta z}-\frac{\dot N}{N}\Big)Q - \Big(\frac{\dot z^3}{N^3\beta z^{\frac{5}{2}}}+\frac{4k\dot z}{N\beta z^{\frac{3}{2}}}\Big)\Lambda(\phi)^2-\Big(\frac{6\dot z^3}{N^3z^{\frac{5}{2}}}+\frac{\alpha \dot{z}}{\beta N\sqrt{z}}+\frac{24k\dot z}{Nz^{\frac{3}{2}}}\Big)\Lambda(\phi)\]
\be \label{p1}p_\phi = \frac{z^{\frac{3}{2}}\dot\phi}{N};\;\;\;\;\;p_N =-\frac{Q\dot z}{N};\;\;\;\;\;p_Q = -\dot z
\ee
Therefore, the Hamiltonian in terms of the phase space variables reads
\be\begin{split}\label{L0}
 H_H &=-p_Q p_z+\frac{N^3\sqrt{z}Q^2}{36\beta}-\Big(\frac{\Lambda(\phi)\dot z^2}{6\beta z}+\frac{N^2\alpha z}{6\beta}+\frac{2kN^2\Lambda(\phi)}{3\beta }+2kN^2\Big)Q\\&
 +\Big(\frac{3\dot z^4}{2N^3z^{\frac{5}{2}}}+\frac{\alpha \dot z^2 }{2\beta N\sqrt{z}}+\frac{2k\alpha N\sqrt {z}}{\beta}+\frac{12k\dot z^2}{Nz^{\frac{3}{2}}}+\frac{24k^2N}{\sqrt z}\Big)\Lambda(\phi)\\&
 +\Big(\frac{\dot z^4}{4N^3\beta z^{\frac{5}{2}}} +\frac{2k\dot z^2}{N\beta z^{\frac{3}{2}}}+\frac{4N k^2}{\beta \sqrt{z}}\Big)\Lambda(\phi)^2+\frac{N\alpha^2 z^{\frac{3}{2}}}{4\beta}+\frac{Np^2_{\phi}}{2z^{\frac{3}{2}}}+VNz^{\frac{3}{2}}.\end{split}\ee
It is finally required to express the Hamiltonian in terms of basic variables ($z$ and $x = {\dot z\over N}$), instead of auxiliary variable. Now, since, $p_Q =-\dot z=-Nx$ and $Q =\frac{p_x}{N}$, one therefore make the following canonical transformation, replacing
$p_Q$ by $-Nx$ and $Q$ by $\frac{p_x}{N}$ in the Hamiltonian (\ref{L0}), to obtain
\be\begin{split}\label{Hl3}& H_H=N\Big[x p_z+\frac{\sqrt{z}p_x^2}{36\beta}-\left(\frac{\alpha z}{6\beta}+\frac{\Lambda(\phi)x^2}{6\beta z}+\frac{2k\Lambda(\phi)}{3\beta }+2k\right)p_x +\frac{p^2_{\phi}}{2z^{\frac{3}{2}}}+\left(\frac{\Lambda^2(\phi)}{4\beta z^{\frac{5}{2}}} +\frac{3\Lambda(\phi)}{2z^{\frac{5}{2}}}\right)x^4+\frac{\alpha^2 z^{\frac{3}{2}}}{4\beta}\\&+\left(\frac{\alpha \Lambda(\phi)}{2\beta \sqrt{z}}+\frac{12k\Lambda(\phi)}{z^{\frac{3}{2}}} +\frac{2k\Lambda^2(\phi)}{\beta z^{\frac{3}{2}}}\right)x^2+\frac{2\alpha k\Lambda(\phi)\sqrt{z}}{\beta}+\frac{4\Lambda^2(\phi)k^2}{\beta \sqrt{z}}+\frac{24k^2\Lambda(\phi)}{\sqrt z}+Vz^{\frac{3}{2}}\Big] = N{\mathcal{H}_{H}}.\end{split}\ee
One can see that all the Hamiltonian \eqref{ho1}, \eqref{Hl2} and \eqref{Hl3} obtained following different routs are the same. The action \eqref{A12} can now be expressed in the canonical form with respect to the basic variables as,
\be\begin{split}\label{Acl} A& = \int\left(\dot z p_z + \dot x p_x + \dot\phi p_{\phi}- N\mathcal{H}_{H}\right)dt~ d^3 x
 = \int\left(\dot h_{ij} \pi^{ij} + \dot K_{ij}\Pi^{ij}+ \dot\phi p_{\phi} - N\mathcal{H}_H \right)dt~ d^3 x,\end{split}\ee
where, $\pi^{ij}$ and $\Pi^{ij}$ are momenta canonically conjugate to $h_{ij}$ and $K_{ij}$ respectively.

\subsubsection{Quantum counterpart}

Due to diffeomorphic invariance, the Hamiltonian is constrained to vanish. Thus canonical quantization leads to the following modified Wheeler-de-Witt equation,
\be\begin{split}\label{q1}&
\frac{i\hbar}{\sqrt z}\frac{\partial \Psi}{\partial z} = -\frac{\hbar^2}{36\beta x}\Big(\frac{\partial^2}{\partial x^2} + \frac{n}{x}\frac{\partial}{\partial x}\Big)\Psi\\&+\frac{i\hbar}{2}\Big(\frac{2k}{\sqrt z}+\frac{\alpha \sqrt z}{6\beta }+\frac{2k\Lambda(\phi)}{3\beta\sqrt z}\Big)\Big(\frac{2}{x}\frac{\partial\Psi}{\partial x}-\frac{\Psi}{x^2}\Big)+\frac{i\hbar\Lambda(\phi)}{12\beta z^{\frac{3}{2}}}\Big(x\frac{\partial\Psi}{\partial x}+\frac{\partial}{\partial x}(x\Psi)\Big)-\frac{\hbar^2}{2xz^2}\frac{\partial^2\Psi}{\partial \phi^2}\\&+\Big(\frac{\alpha^2z}{4\beta x}+\frac{\Lambda^2x^3}{4\beta z^3}+\frac{4\Lambda^2k^2}{\beta x z}+\frac{\alpha \Lambda x }{2\beta {z}}+\frac{2k\alpha\Lambda}{\beta x}+\frac{12k\Lambda x}{z}+\frac{2k\Lambda^2x}{\beta z^{2}}+\frac{24k^2\Lambda}{xz}+\frac{3\Lambda x^3}{2z^{2}}+\frac{V z}{x}\Big)\Psi.\end{split}\ee
Where Weyl symmetric ordering has been performed in the 1st., 2nd. and 3rd. term appearing on right hand side and $n$ is the operator ordering index. Now, again under a further change of variable, the above modified Wheeler-de-Witt equation, takes the look of Schrodinger equation, viz.,
 \be\begin{split}\label{q12}&
i\hbar\frac{\partial \Psi}{\partial \sigma} = -\frac{\hbar^2}{54\beta x}\left(\frac{\partial^2}{\partial x^2} + \frac{n}{x}\frac{\partial}{\partial x}\right)\Psi+i\hbar\left(\frac{2k}{3\sigma^{\frac{1}{3}}}+\frac{\alpha \sigma^{\frac{1}{3}}}{18\beta }+\frac{4k\Lambda(\phi)}{9\beta \sigma^{\frac{1}{3}}}\right)\left(\frac{1}{x}\frac{\partial\Psi}{\partial x}-\frac{\Psi}{2x^2}\right)\\&+\frac{i\hbar\Lambda(\phi)}{18\beta \sigma}\left(2x\frac{\partial\Psi}{\partial x}+\Psi\right)-\frac{\hbar^2}{2x\sigma^{\frac{4}{3}}}\frac{\partial^2\Psi}{\partial \phi^2}+V_e\Psi=\hat{H_e}\Psi.\end{split}\ee
Where $\sigma=z^{\frac{3}{2}}=a^3$ plays the role of internal time parameter, as before. In the above expression \eqref{q12}, the effective potential $V_e$, is given by, \be V_e=\frac{\alpha^2\sigma^{\frac{2}{3}}}{6\beta x}+\Big(\frac{x^3}{6\beta\sigma^2}+\frac{4k x}{3\beta \sigma^{\frac{4}{3}}}+\frac{8k^2}{3\beta x \sigma^{\frac{2}{3}}}\Big)\Lambda(\phi)^2+
\Big(\frac{x^3}{\sigma^{\frac{4}{3}}}+\frac{\alpha x }{3\beta {\sigma^{\frac{2}{3}}}}+\frac{4k\alpha}{3\beta x}+\frac{8kx}{\sigma^{\frac{2}{3}}}+\frac{16k^2}{x\sigma^{\frac{2}{3}}}
\Big)\Lambda(\phi)\\+\frac{V\sigma^{\frac{2}{3}}}{x}\ee
The hermiticity of the effective Hamiltonian allows one to write the continuity equation for $n = -1$ as,
\be   \frac{\partial\rho}{\partial\sigma}+\nabla.\mathbf{J}=0,\ee
where, $\rho=\Psi^{*}\Psi$ and $\mathbf{J} = (J_x, J_{\phi}, 0)$ are the probability density and the current density respectively, with, $J_x = \frac{i\hbar}{54\beta x}(\Psi^{*}_{,x}\Psi-\Psi^{*}\Psi_{,x})-\left(\frac{2k}{3\sigma^{\frac{1}{3}}}+\frac{\alpha\sigma^{\frac{1}{3}} }{18\beta }\right)\frac{\Psi^{*}\Psi}{x}-\frac{\Lambda(\phi) x}{18\beta \sigma}\Psi^{*}\Psi$ and $J_{\phi}=\frac{i\hbar}{2x\sigma^{\frac{4}{3}}}(\Psi^{*}_{,\phi}\Psi-\Psi^{*}\Psi_{,\phi})$.

\subsubsection{Semiclassical solution under WKB approximation}

Now to check the viability of the quantum equation (\ref{q12}), it is required to test its behaviour under certain appropriate semi-classical approximation. The semi-classical approximation may be performed only about a known classical solution of the field equations. To obtain an appreciable solution, let us fix the gauge $N = 1$, and set $k = 0$, so that the $\phi$ variation equation \eqref{phi} and the $(^0_0)$ component of Einstein's equation \eqref{A22} are expressed respectively in terms of the scale factor $(a)$ as

\be - 24\Lambda'\dot a^2\ddot a + a^3\left(\ddot\phi + 3{\dot a\over a}\dot\phi +  V'\right) = 0.\ee
\be\begin{split}\label{clfld5}
\frac{\dot a^2}{a^2} &=  - \frac{6\beta}{\alpha}\left[2\frac{\dot a\dddot a}{a^2} - \frac{\ddot a^2}{a^2} + 2\frac{\dot a^2\ddot a}{a^3} - 3\frac{\dot a^4}{a^4} \right] - \frac{4 \Lambda' \dot a\dot\phi}{\alpha} \left(\frac{\dot a^2}{a^3}\right) + \frac{1}{6\alpha}\left(\frac{\dot\phi^2}{2} + V\right).
\end{split}\ee
The above set of equations admit the following inflationary solutions,
\be\begin{split}\label{111}&a=a_0e^{\mathrm{H}t}~~\text{and} ~~\phi=\phi_0 e^{-\mathrm{H}t}\\&
\mathrm{under ~the~ condition,}\;\; \Lambda(\phi) = -{\phi^2\over 48\mathrm{H}^2};~~
V = {6\alpha\mathrm{H}^2} +{1\over 2}\mathrm{H}^2\phi^2.\end{split}\ee
Now, to perform the semiclassical approximation, let us express equation (\ref{q1}) in the form,
\be\begin{split}\label{11}&
 -\frac{\hbar^2\sqrt z}{36\beta x}\left(\frac{\partial^2}{\partial x^2} + \frac{n}{x}\frac{\partial}{\partial x}\right)\Psi -\frac{\hbar^2}{2xz^{\frac{3}{2}}}\frac{\partial^2\Psi}{\partial \phi^2}-{i\hbar}\frac{\partial \Psi}{\partial z}+\frac{i\hbar\alpha z}{6\beta x}\frac{\partial\Psi}{\partial x}-\frac{i\hbar \phi^2 x}{288\beta \mathrm{H}^2 z}\frac{\partial\Psi}{\partial x}+\mathcal V\Psi=0,\\&
\mathrm{where},\;\;\mathcal V=\left(\frac{\alpha^2z^{\frac{3}{2}}}{4\beta x}+\frac{\phi^4x^3}{9216\beta \mathrm{H}^4 z^{\frac{5}{2}}}-\frac{\alpha \phi^2x}{96 \beta \mathrm{H}^2{\sqrt z}}-\frac{\phi^2x^3}{32\mathrm{H}^2z^{\frac{3}{2}}}-\frac{i\hbar\alpha z}{12\beta x^{2}} -\frac{i\hbar \phi^2}{576\beta \mathrm{H}^2 z}+\frac{6\alpha\mathrm{H}^2z^{\frac{3}{2}}}{x} + \frac{\mathrm{H}^2\phi^2z^{\frac{3}{2}}}{2 x}\right).\end{split}\ee
The above equation may be treated as time independent Schr\"{o}dinger equation with three variables $x$, $z$ and $\phi$. Hence, as usual let us seek the solution of the wave-equation (\ref{11}) as,
\be\label{psi} \Psi=\Psi_0 e^{\frac{i}{\hbar}S(x,z,\phi)}\ee
and expand $S$ in power series of $\hbar$ as,
\be\label{S1} S = S_0(x, z,\phi) + \hbar S_1(x, z, \phi) + \hbar^2S_2(x, z, \phi) + .... .\ee
Now, inserting the expressions (\ref{psi}) and (\ref{S1}) in equation (\ref{11}) and equating the coefficients of different powers of $\hbar$ to zero, the following set of equations (upto second order) are obtained.
\begin{subequations}\begin{align}
&\label{s0}\frac{\sqrt z}{36\beta x}S_{0,x}^2+\frac{S_{0,\phi}^2}{2xz^{\frac{3}{2}}}+S_{0,z}-\frac{\alpha z}{6\beta x}S_{0,x} +\frac{\phi^2 x} {288\beta \mathrm{H}^2 z}S_{0,x} + {\mathcal V}(x,z,\phi)= 0\\
&\label{q11}-\frac{i\sqrt z}{36\beta x}S_{0,xx}-\frac{in\sqrt z}{36\beta x^2}S_{0,x}-\frac{iS_{0,\phi\phi}}{2xz^{\frac{3}{2}}}+S_{1,z}+\frac{\sqrt zS_{0,x}S_{1,x}}{18\beta x}+\frac{S_{0,\phi}S_{1,\phi}}{xz^{\frac{3}{2}}}-\frac{\alpha z}{6\beta x}S_{1,x}+\frac{\phi^2 x} {288\beta \mathrm{H}^2 z}S_{1,x}=0\\
&-\frac{i\sqrt z}{36\beta x}S_{1,xx}-\frac{in\sqrt z}{36\beta x^2}S_{1,x}-\frac{iS_{1,\phi\phi}}{2xz^{\frac{3}{2}}}+S_{2,z}+\frac{\sqrt zS_{0,x}S_{2,x}}{18\beta x}+\frac{S_{0,\phi}S_{2,\phi}}{xz^{\frac{3}{2}}}-\frac{\alpha z}{6\beta x}S_{2,x}+\frac{\phi^2 x} {288\beta \mathrm{H}^2 z}S_{2,x}=0
\end{align}\end{subequations}
which are to be solved successively to find $S_0(x, z,\phi)$, $S_1(x, z,\phi)$ and $S_2(x, z,\phi)$ and so on. Now identifying $S_{0,x}$ with $p_x$, $S_{0,z}$ with $p_z$ and $S_{0,\phi}$ with $p_\phi$, and in view of equation (\ref{s0}), one can recover the classical ($^0_0$) component of Einstein's equation, presented in (\ref{A22}). Thus, $S_0(x, z)$ can now be expressed as,
\be\label{S} S_0=\int p_z dz+\int p_x dx+\int p_\phi d\phi\ee
apart from a constant of integration which may be absorbed in $\Psi_0$. The integrals in the above expression can be
evaluated using the classical solution for $k = 0$ presented in equation (\ref{111}), the definition of $p_z$ and $p_\phi$ presented
in (\ref{p1}) and $p_x = NQ$. Recalling the expression for $Q$ given in (\ref{Q}) together with the relation, $x = \dot z$, where, $z = a^2$, and choosing the value of operator ordering index $n = -1$, for which probability interpretation holds, one can use solution (\ref{111}), to express $x$, $p_x$, $p_z$ and $p_\phi$ in terms of $z$ and $\phi$ as,
\begin{subequations}\begin{align}
&x=2\mathrm{H}z\\
&p_x=Q={\frac{3\alpha \sqrt x}{\sqrt {2\mathrm{H}}}} +\frac{72\beta \mathrm{H}^{\frac{3}{2}}\sqrt x}{\sqrt {2}}-\frac{\sqrt{\mathrm{H}}a_0^2\phi_0^2}{2\sqrt2\sqrt x}\\
&p_z=-3\alpha\mathrm{H}\sqrt z-72\beta \mathrm{H}^3\sqrt z-\frac{\mathrm{H} a_0^2\phi_0^2}{4\sqrt z}\\
&p_\phi=-\frac{\mathrm{H}a_0^3\phi_0^3}{\phi^2}
\end{align}\end{subequations}
Hence the integrals in (\ref{S}) are evaluated as,
\begin{subequations}\begin{align}
&\int p_x dx=\frac{2\alpha}{\sqrt {2\mathrm{H}}}x^{\frac{3}{2}}+\frac{48\beta \mathrm{H}^{3\over 2}}{\sqrt 2} x^{\frac{3}{2}}-{\sqrt {2 \mathrm{H}} a_0^2\phi_0^2\over 2}\sqrt {{x}} = 4\alpha\mathrm{H} z^{3\over 2} + 96\beta \mathrm{H}^3 z^{3\over 2} - a_0^2\phi_0^2 \mathrm{H}\sqrt z \\
&\int p_z dz=-2\alpha \mathrm{H}z^{\frac{3}{2}}-48\beta \mathrm{H}^3z^{\frac{3}{2}}-{\mathrm{H} a_0^2\phi_0^2\over 2}{\sqrt z}\\
&\int p_\phi d\phi=\frac{\mathrm{H}a_0^3\phi_0^3}{\phi} = \mathrm{H}a_0^2\phi_0^2\sqrt z
\end{align}\end{subequations}
Therefore, the explicit form of the Hamilton-Jacobi function $S_0$ in terms of z is found as,
\be\label{S01} S_0=2\alpha \mathrm{H} z^{\frac{3}{2}}+48\beta \mathrm{H}^3z^{\frac{3}{2}} -{\mathrm{H}a_0^2\phi_0^2\over 2}\sqrt z\ee
In addition, one can also compute the zeroth order on-shell action using classical solution \eqref{111} in action \eqref{A12} or equivalently in (\ref{71}), and express all the variables in terms of $t$ to obtain
\be\label{2.a} A = A_{cl} = \int\Bigg[ 6\alpha \mathrm{H}^2 {a_0^3} e^{3 \mathrm{H} t}+144 \beta \mathrm{H}^4{a_0^3}e^{3 \mathrm{H} t} -\frac{\mathrm{H}^2 {a_0^3} {\phi_0^2} e^{\mathrm{H} t}}{2}\Bigg]dt.\ee
On integration, one thus finds the zeroth order on-shell action as,
\be\label{A} A = A_{cl} = 2\alpha \mathrm{H} {a_0^3} e^{3 \mathrm{H} t}+48\beta \mathrm{H}^3{a_0^3}e^{3 \mathrm{H} t} -\frac{\mathrm{H} {a_0^3} {\phi_0^2} e^{\mathrm{H} t}}{2},\ee
which is the same as the Hamilton-Jacobi function $S_0$ obtained in equation \eqref{S01}. Note that same result would have been obtained if one uses the canonical action \eqref{Acl} instead. Since everything is fair, so one can compute the semiclassical wave function, which now reads as,

\be \Psi = \psi_{0} e^{{i\over \hbar}[2\alpha \mathrm{H} z^{\frac{3}{2}}+48\beta \mathrm{H}^3z^{\frac{3}{2}} -{\mathrm{H}a_0^2\phi_0^2\over 2}\sqrt z]}.\ee
Further, using the expression for $S_0$ obtained in \eqref{S01}, it is also possible to find the expression for $S_{1,z}$in view of \eqref{q11}, which upon integration yields,

\be S_1 = i g(z).\ee
Thus, the semi-classical wavefunction upto first order approximation reads

\be\label{psid} \Psi = \psi_{01} e^{{1\over\hbar} [2\alpha \mathrm{H} z^{\frac{3}{2}}+48\beta \mathrm{H}^3z^{\frac{3}{2}} -{\mathrm{H}a_0^2\phi_0^2\over 2}\sqrt z]}, ~~\mathrm{where},~~ \psi_{01} = \psi_0 e^{g(z)}.\ee
The oscillatory behaviour of the wavefunction \eqref{psid} indicates that the region is classically allowed and the wavefunction is strongly peaked about a set of exponential solutions \eqref{111} to the classical field equations. This establishes the correspondence between the quantum equation and the classical equations. So, everything is consistent in all the three formalisms handled so far.

\subsection{Modified Horowitz' formalism}

In this subsection we invoke modified Horowitz' canonical formulation, to find that it also produces a phase-space Hamiltonian which results in a viable quantum dynamics. Nevertheless, what we find is a different Hamiltonian altogether, which is not canonically related to the earlier ones. We follow the same route as presented in subsection (2.4), i.e. split the boundary term associated with $R^2$ into two parts \eqref{sigma}, integrate action (\ref{A12}) so that all total derivative terms get cancelled with the supplementary boundary terms. Therefore, the action to start with in the Modified Horowitz' formalism is the following,

\be\begin{split}\label{15.1}
A &= \int\Bigg[\alpha\left( - \frac{3 {\dot z}^2}{2 N \sqrt z} + 6 k N \sqrt z \right) + \frac{9 \beta}{\sqrt z}\left(\frac{{\ddot z}^2}{N^3} - \frac{2 \dot N \dot z \ddot z}{N^4} + \frac{{\dot N}^2{\dot z}^2}{N^5} + \frac{2 k {\dot z}^2}{N z} + 4 k^2 N \right) \\&\hspace{1.0 in}- \frac {\Lambda^{'} \dot z \dot {\phi}}{N \sqrt z}\left(\frac{{\dot z}^2}{N^2 z} + 12 k\right)+z^{\frac{3}{2}}\left(\frac{1}{2N}\dot\phi^2-VN\right)\Bigg]dt + \Sigma_{R^2_2},
\end{split}\ee
instead of \eqref{A12}. The auxiliary variable
\be\label{al} \mathcal Q=\frac{18 \beta}{N^3\sqrt z}\left({\ddot z}-\frac{\dot N \dot z}{ N}\right)\ee
is introduced at this stage straight into the action (\ref{15.1}) as,
\be\begin{split}\label{Lt2}
A &= \int\Bigg[\mathcal Q\ddot z-\frac{N^3\sqrt{z}}{36\beta}{\mathcal Q}^2-\frac{\dot z\dot N \mathcal Q}{N} -\frac{3\alpha\dot z^2}{2N\sqrt z}+6\alpha kN\sqrt{z}+ \frac{18\beta k\dot z^2}{Nz^{\frac{3}{2}}} +\frac{36\beta Nk^2}{\sqrt{z}} -\frac{\Lambda'(\phi)\dot\phi{\dot z}^3}{z^{\frac{3}{2}}}\\& \hspace{1.0 in} -12k\frac{\Lambda'(\phi)\dot\phi\dot z}{\sqrt{z}}+z^{\frac{3}{2}}\left(\frac{1}{2N}\dot\phi^2-VN\right)\Bigg]dt+ \Sigma_{R^2_2},\end{split}\ee
and under integration by parts, the rest of the supplementary boundary terms is taken care of. The canonical action therefore takes the following form,

\be\begin{split}\label{Lt23}
A &= \int\Bigg[-\dot {\mathcal Q}\dot z-\frac{N^3\sqrt{z}}{36\beta}{\mathcal Q}^2-\frac{\dot z\dot N \mathcal Q}{N} -\frac{3\alpha\dot z^2}{2N\sqrt z}+6\alpha kN\sqrt{z}+ \frac{18\beta k\dot z^2}{Nz^{\frac{3}{2}}} +\frac{36\beta Nk^2}{\sqrt{z}} -\frac{\Lambda'(\phi)\dot\phi{\dot z}^3}{N^3z^{\frac{3}{2}}}\\&\hspace{1.0 in} -\frac{12k\Lambda'(\phi)\dot\phi\dot z}{N\sqrt{z}}+z^{\frac{3}{2}}\left(\frac{1}{2N}\dot\phi^2-VN\right)\Bigg]dt.\end{split}\ee
The canonical momenta are
\[
p_{\mathcal Q} = - \dot z;\;\;\;\;\;p_z =-\dot {\mathcal Q} - \frac{\mathcal Q \dot N}{N}-\frac{3\alpha\dot z}{N\sqrt{z}}+\frac{36 \beta k \dot z}{N z^{\frac {3}{2}}}- \frac{3\Lambda'(\phi)\dot\phi\dot z^2}{N^3z^{\frac{3}{2}}} -\frac{12k\Lambda'(\phi)\dot\phi}{N\sqrt{z}}\label{12};\]
\be \label{mom}p_\phi = - \frac{\Lambda' \dot z}{N \sqrt z}\left(\frac{\dot z^2}{N^2 z} + 12 k\right)+\frac{z^{\frac{3}{2}}}{N}\dot\phi;\;\;\;\;p_N = -\frac{\mathcal Q \dot z}{N}
\ee
Now, in view of the above definitions of momenta \eqref{mom} one can find the following relation,

\be p_{\mathcal Q} p_z = \frac{3\alpha\dot z^2}{ N \sqrt z} + \dot z\dot{\mathcal Q}+ \frac{\dot N}{N}\dot z \mathcal Q - \frac{36 k \beta \dot z^2}{N z^{\frac{3}{2}}} + \frac{3 \dot z \Lambda^{'} \dot\phi}{N \sqrt z}\left(\frac{\dot z^2}{N^2 z} + 4k\right)\ee
which is used to find the phase space structure of the Hamiltonian as,

\be\begin{split}\label{2}
H &= 3\alpha\Big(\frac{{p_{\mathcal Q}}^2}{2 N \sqrt z} - 2k N \sqrt z \Big) - p_{\mathcal Q} p_z + \frac{N^3 {\mathcal Q}^2 \sqrt z}{36\beta} - \frac{18 k \beta}{\sqrt z} \Big(\frac{{p_\mathcal Q}^2}{N z} + 2k N\Big) +\frac{Np_{\phi}^2}{2z^{\frac{3}{2}}}\\&\hspace{0.5in}+\frac{\Lambda'^2p_Q^6}{2N^5z^{\frac{9}{2}}}+\frac{12k\Lambda'^2p_Q^4}{N^3z^{\frac{7}{2}}}+ \frac{72k^2\Lambda'^2p_Q^2}{Nz^{\frac{5}{2}}}-\frac{\Lambda'p_Q^3p_{\phi}}{N^2z^3}-\frac{12k\Lambda'p_Qp_{\phi}}{z^2}+NVz^{\frac{3}{2}}.
\end{split}\ee
Finally, under the choice, $x = \frac{\dot z}{N}$, the canonical transformations, ${\mathcal Q} = \frac{p_x}{N}$ and $p_{\mathcal Q} = -\dot z = -Nx$, results in the Hamiltonian in terms of the basic variables as,

\be\begin{split}\label{3}
H &= N\Bigg[x p_z + \frac{ \sqrt z {p_x}^2}{36\beta} +\frac{p_{\phi}^2}{2z^{\frac{3}{2}}}+\left(\frac{x^3}{z^3}+\frac{12kx}{z^2}\right)\Lambda'p_{\phi}+ 3\alpha\Big(\frac{x^2}{2 \sqrt z} - 2k \sqrt z \Big)\\& \hspace{0.50 in}- \frac{18 k \beta}{\sqrt z} \Big(\frac{x^2}{z} + 2k\Big)+\left(\frac{x^6}{2z^{\frac{9}{2}}}+\frac{12kx^4}{z^{\frac{7}{2}}}+ \frac{72k^2x^2}{z^{\frac{5}{2}}}\right)\Lambda'^2+Vz^{\frac{3}{2}} \Bigg]= N{\mathcal H}.
\end{split}\ee
One can clearly observe the difference between the Hamiltonian \eqref{3} with the one obtained following earlier techniques (\ref{Hl3}). While, the present one involves derivative of the coupling parameter ($\Lambda'$), earlier ones didn't. Action (\ref{15.1}) can now be expressed (instead of action \eqref{A12}) in the canonical form with respect to the basic variables as,

\be\begin{split}\label{ACL} A &= \int\left(\dot z p_z + \dot x p_x - N\mathcal{H}_L\right)dt~ d^3 x =
\int\left(\dot h_{ij} \pi^{ij} + \dot K_{ij}\Pi^{ij} - N\mathcal{H}_L\right)dt~ d^3 x,\end{split}\ee
where, $\pi^{ij}$ and $\Pi^{ij}$ are momenta canonically conjugate to $h_{ij}$ and $K_{ij}$ respectively. As in the previous case encountered in section (2), here again we observe that the phase-space Hamiltonian obtained under modification of Horowitz' formalism, is different from the ones (\ref{ho1}/\ref{Hl2}/\ref{Hl3}) obtained following standard techniques. One can try to find a transformation here again as,
\begin{multline}\begin{split}\label{tr}
  z=Z,&\; p_z= P_Z-18\beta\frac{kX}{Z^{\frac{3}{2}}}+\frac{3\alpha X}{2\sqrt{Z}}-\frac{6k\Lambda X}{Z^{\frac{3}{2}}}-\frac{\Lambda X^3}{2Z^{\frac{5}{2}}};\\ x=X,&\;p_x=P_X+36\beta \frac{k}{\sqrt{Z}}+3\alpha\sqrt{Z}-\frac{3\Lambda X^2}{Z^{\frac{3}{2}}}-\frac{12k\Lambda}{\sqrt Z};\phi=\Phi,\;p_{\phi}=P_{\Phi}+\frac{\Lambda' X^3}{Z^{\frac{3}{2}}}+\frac{12k\Lambda' X}{\sqrt Z},
\end{split}\end{multline}
to show that The Hamiltonian are one and the same. But unfortunately the transformations are not canonical \cite{ourpu}. One can further note that canonical quantization in the present case \eqref{3} requires operator ordering between $\Lambda(\phi)$ and $p_{\phi}$, which depends on the specific form of $\Lambda$. This situation didn't arise in earlier canonical formulations (\ref{ho1}/\ref{Hl2}/\ref{Hl3}). In this respect again the latter canonical version differs from the earlier ones. Thus, we prove that choice of boundary term indeed tells upon the canonical structure of the theory.

\subsubsection{Canonical Quantization}

We have obtained a different phase-space Hamiltonian \eqref{3}, fixing $\delta h_{ij} = 0 = \delta R$ at the boundary. Already we have observed that either of the standard canonical formulation techniques (Ostrogradski's/ Dirac's/ Horowitz') does produce a viable Hamiltonian of the theory. Now naturally the question - ``Is the Hamiltonian and its quantum counterpart obtained following modified Horowitz' technique viable?" must be answered. The accountability of the Hamiltonian \eqref{3} may be explored through a viable semiclassical approximation corresponding to the modified Wheeler-de-Witt equation, which reads,
\be\begin{split}\label{7.1}
\frac{i\hbar}{\sqrt z}\frac{\partial \Psi}{\partial z} &= -\frac{\hbar^2}{36\beta x}\left(\frac{\partial^2}{\partial x^2} + \frac{n}{x}\frac{\partial}{\partial x}\right)\Psi -\frac{\hbar^2}{2xz^2}\frac{\partial^2 \Psi}{\partial \phi^2}-\frac{1}{2z^{\frac{5}{2}}}\left(\frac{x^2}{z}+12k\right)\widehat{\Lambda'}\widehat{p_{\phi}}\\&+ \left[3\alpha\left(\frac{x}{2 z} - \frac{2k}{x}\right) - \frac{18 k \beta}{z} \left(\frac{x}{z} +\frac{2k}{x}\right)+\frac{\Lambda'^2x}{z^3}\left(\frac{x^4}{2z^2}+\frac{12kx^2}{z}+72k^2\right)+\frac{Vz^{\frac{3}{2}}}{x}\right]\Psi = \hat H_e\Psi.
\end{split}\ee
where, $n$ is the operator ordering index. Operator form of $\widehat\Lambda' \widehat {p_{\phi}}$ appearing on the third term on the right hand side may be performed only after having knowledge of a specific form of $\Lambda(\phi)$. Specific form of $\Lambda(\phi)$ is also required to investigate the behaviour of the quantum theory, under certain appropriate semi-classical approximation, which may only be obtained from the solution of the classical field equations \eqref{phi} and \eqref{A22}.  We use the same inflationary solutions presented in \eqref{111}. Correspondingly, the modified Wheeler-de-Witt equation (\ref{7.1}) may now be expressed as

\be\begin{split}\label{7}
\frac{i\hbar}{\sqrt z}\frac{\partial \Psi}{\partial z} &= -\frac{\hbar^2}{36\beta x}\Big(\frac{\partial^2}{\partial x^2} + \frac{n}{x}\frac{\partial}{\partial x}\Big)\Psi - \frac{\hbar^2}{2x z^2}\frac{\partial^2\Psi}{\partial \phi^2} + i\hbar \frac{x^2}{48 \mathrm{H}^2z^{\frac{7}{2}}}\Big(\phi\frac{\partial \Psi}{\partial \phi} + \frac{\partial}{\partial \phi}(\phi\Psi)\Big)\\& \hspace{1.45 in}+ \left(\frac{3\alpha x}{2z} + \frac{\phi^2 x^5}{576 \mathrm{H}^4 z^5} +\Big(6\alpha \mathrm{H}^2 + {1\over 2}\mathrm{H}^2\phi^2\Big) {z^{\frac{3}{2}}\over x}\right)\Psi,\end{split}\ee
where Weyl symmetric ordering has been performed in the third term appearing on the right hand side. Now, again under a further change of variable, the above modified Wheeler-de-Witt equation, takes the look of Schr\"odinger equation, viz.,

\be\begin{split}\label{5}
i\hbar\frac{\partial \Psi}{\partial \sigma} &= -\frac{\hbar^2}{54\beta}\left(\frac{1}{x}\frac{\partial^2}{\partial x^2} + \frac{n}{x^2}\frac{\partial}{\partial x}\right)\Psi - \frac{\hbar^2}{3x \sigma^{\frac{4}{3}}}\frac{\partial^2\Psi}{\partial \phi^2} + i\hbar \frac{x^2}{72 \mathrm{H}^2\sigma^{\frac{7}{3}}}\left(2\phi\frac{\partial \Psi}{\partial \phi} + \Psi\right) + V_e\Psi = \hat H_e\Psi,
\end{split}\ee
where, $\sigma  = z^{\frac{3}{2}} = a^3$ plays the role of internal time parameter as before. In the above, the effective potential $V_e$, is given by,
\be\label{Ve} V_e = \frac{\alpha x}{\sigma^{\frac{2}{3}}} + \frac{\phi^2 x^5}{864 \mathrm{H}^4\sigma^{\frac{10}{3}}} + \frac{2\sigma}{3x}\Big(6\alpha \mathrm{H}^2 +{1\over 2}\mathrm{H}^2\phi^2\Big)  \ee
The hermiticity of $\hat H_e$ allows one to write the continuity equation under the choice $n = -1$, as,
\be \frac{\partial\rho}{\partial \sigma} + \nabla . \mathbf{J} = 0, \ee
where, $ \rho = \Psi^*\Psi ~~ \text{and} ~~  \mathbf{J} = (J_x, J_\phi, 0) $ are the probability density and the current density respectively, where
\begin{subequations}\begin{align}
{J}_x &= \frac{i \hbar }{54\beta x}(\Psi\Psi^*_{,x}-\Psi^*\Psi_{~,x})\\
{J}_\phi &= \frac{i \hbar }{3x\sigma^{\frac{4}{3}}}(\Psi\Psi^*_{,\phi}-\Psi^*\Psi_{~,\phi}) - \frac{x^2\phi}{36 \mathrm{H}^2\sigma^{\frac{7}{3}}}\Psi^*\Psi
\end{align}
\end{subequations}
In the process, operator ordering index here too has been fixed as $n = -1$ from physical argument.

\subsubsection{Semiclassical approximation}

Now to check the viability of the quantum equation (\ref{5}), it is required to test its behaviour under certain appropriate semi-classical approximation. For the purpose, let us express equation (\ref{7}) in the form,
\be\begin{split}\label{8}
-\frac{\hbar^2\sqrt z}{36\beta x}\left(\frac{\partial^2}{\partial x^2} + \frac{n}{x}\frac{\partial}{\partial x}\right)\Psi - \frac{\hbar^2}{2x z^{\frac{3}{2}}}\frac{\partial^2\Psi}{\partial \phi^2} - i\hbar\frac{\partial \Psi}{\partial z} + i\hbar \frac{x^2}{24 \mathrm{H}^2 z^3}\phi\frac{\partial \Psi}{\partial \phi} + \mathcal{V}\Psi = 0
\end{split}\ee
where
\be \mathcal{V} = \frac{3\alpha x}{2\sqrt{z}} + \frac{\phi^2 x^5}{576 \mathrm{H}^4 z^{\frac{9}{2}}} + \frac{i\hbar x^2}{48 \mathrm{H}^2z^3} + \Big(6\alpha\mathrm{H}^2 + {1\over 2}\mathrm{H}^2\phi^2\Big)\frac{z^2}{x}.\ee
The above equation may be treated as time independent Schr{\"o}dinger equation with three variables $x$, $z$ and $\phi$. Therefore, as usual, let us seek the solution of equation (\ref{8}) as,
\be\label{9} \psi = \psi_0e^{\frac{i}{\hbar}S(x,z,\phi)}\ee
and expand $S$ in power series of $\hbar$ as,
\be\label{10} S = S_0(x,z,\phi) + \hbar S_1(x,z,\phi) + \hbar^2S_2(x,z,\phi) + .... \ .\ee
Now inserting the expressions (\ref{9}) and (\ref{10}) in equation (\ref{8}) and equating the coefficients of different powers of $\hbar$ to zero, one obtains the following set of equations (upto second order)
\begin{subequations}\begin{align}
&\frac{\sqrt z}{36\beta x}S_{0,x}^2 + \frac{S_{0,\phi}^2}{2xz^{\frac{3}{2}}} + S_{0,z} - \frac{x^2}{48 \mathrm{H}^2 z^3}\phi S_{0,\phi} +  \mathcal{V}(x,z,\phi) = 0 \label{hbar0} \\
& -\frac{i\sqrt z}{36\beta x}S_{0,xx} - \frac{in\sqrt z}{36\beta x^2}S_{0,x} - \frac{iS_{0,\phi\phi}}{2xz^{\frac{3}{2}}} + S_{1,z} + \frac{\sqrt zS_{0,x}S_{1,x}}{18\beta x} + \frac{S_{0,\phi}S_{1,\phi}}{xz^{\frac{3}{2}}} - \frac{x^2}{48 \mathrm{H}^2 z^3}\phi S_{1,\phi} = 0. \label{hbar1}\\
& \frac{\sqrt z S_{0,x}S_{2,x}}{18\beta x} - i\frac{\sqrt z S_{1,xx}}{36\beta x} - i\frac{n\sqrt zS_{1,x}}{36\beta x^2} + \frac{S_{0,\phi}S_{2,\phi}}{xz^{\frac{3}{2}}} - i\frac{S_{1,\phi\phi}}{xz^{\frac{3}{2}}} + S_{2,z} - \frac{x^2}{48 \mathrm{H}^2 z^3}\phi S_{2,\phi} = 0,
\end{align}\end{subequations}
which are to be solved successively to find $S_0(x,z,\phi),\; S_1(x,z,\phi)$ and $S_2(x,z,\phi)$ and so on. Now identifying $S_{0,x}$ with $p_x$; $S_{0,z}$ with $p_z$ and $S_{0,\phi}$ with $p_{\phi}$ one can recover the classical Hamiltonian constraint equation $H = 0$, given in equation (\ref{3}) from equation (\ref{hbar0}). Thus, $S_{0}(x, z)$ can now be expressed as,

\be\label{14} S_0 = \int p_z dz + \int p_x dx + \int p_\phi d\phi \ee
apart from a constant of integration which may be absorbed in $\psi_0$. The integrals in the above expression may be evaluated using the classical solution for $k = 0$, presented in equation (\ref{111}), the definition of $p_z$, $p_{\phi}$ in (\ref{mom}) and $p_x = N Q$. Further, recalling the expression for $Q$ given in (\ref{al}), the relation, $x = \dot z$, where, $z = a^2$, the choice $n = -1$, for which probability interpretation holds, and using the solution (\ref{111}), $x$, $p_x$, $p_z$, and $p_\phi$ are found as,
\begin{subequations}\begin{align}
&\label{comb}
x = 2{\mathrm H} z \\
&p_x = 36\sqrt 2 \beta {\mathrm H}^{\frac{3}{2}}\sqrt {{x}}  \\
&p_z = -{6\alpha}{\mathrm H}\sqrt z - 72\beta {\mathrm H}^3\sqrt z  -{{\mathrm H} a_0^2\phi_0^2\over 2\sqrt z} \\
&p_\phi = -\frac{ 2{\mathrm H}a_0^3\phi_0^3}{3\phi^2}
\end{align}\end{subequations}
Hence the integrals in (\ref{14}) are evaluated as,
\begin{subequations}\begin{align}
&\int p_x dx = 24\sqrt 2 \beta {\mathrm H}^{\frac{3}{2}} {x^{\frac{3}{2}}} = 96\beta {\mathrm H}^3 z^{3\over 2}; \\
&\int p_z dz =  -4{\alpha}\mathrm{H} z^{\frac{3}{2}} - 48\beta\mathrm{H}^3 z^{\frac{3}{2}} - {\mathrm H}a_0^2\phi_0^2 \sqrt z; \\
&\int p_\phi d\phi = \frac{2{\mathrm H}a_0^3\phi_0^3}{3\phi} ={2 \over 3}{\mathrm H}a_0^2\phi_0^2\sqrt z.
\end{align}\end{subequations}
\noindent
and the explicit form of $S_0$ in terms of $z$ can be written as,
\be\label{S0} S_0 = -{4\alpha}\mathrm{H} z^{\frac{3}{2}} + 48\beta\mathrm{H}^3 z^{\frac{3}{2}} - {\mathrm H\over 3}a_0^2\phi_0^2\sqrt z.\ee
So, for consistency, one can trivially check that the expression for $S_0$ (\ref{S0}) so obtained, satisfies equation (\ref{hbar0}) identically. In fact it should, because, equation (\ref{hbar0}) coincides with Hamiltonian constraint equation (\ref{3}) for $k = 0$. Moreover, one can also compute the zeroth order on-shell action. It is very important to mention that one should substitute classical solution (\ref{111}) in the action we started with in `Modified Horowitz Formalism', viz. \eqref{15.1} or equivalently in (\ref{Lt2}), instead of \eqref{A12}. One may then express all the variables in terms of $t$ to obtain

\be\label{A01} A=A_{cl}=\int\left[-{12\alpha}{\mathrm{H}}^2 a_0^3 e^{3\mathrm{H}t} + 144\beta\mathrm{H}^4 a_0^3 e^{3\mathrm{H}t} - {\mathrm{H}^2\over 3} a_0^3\phi_0^2 e^{\mathrm{H}t}\right]dt.\ee
Integrating we have, \be \label{AS0} A=A_{cl}=-{4\alpha}{\mathrm{H}} a_0^3 e^{3\mathrm{H}t} + 48\beta\mathrm{H}^3 a_0^3 e^{3\mathrm{H}t} - {\mathrm{H}\over 3}a_0^3\phi_0^2 e^{\mathrm{H}t},\ee
which is the same as we obtained in (\ref{S0}). Same result would have been found in view of the canonical action \eqref{ACL}. Since there is no inconsistency found at any stage, so it is clear that the modified Horowitz' canonical formulation is also on the right track. At this end, the wave function takes the form,

\be \psi = \psi_0 e^{\frac{i}{\hbar}\left[-4{\alpha}\mathrm{H} z^{\frac{3}{2}} + 48\beta\mathrm{H}^3 z^{\frac{3}{2}} -{\mathrm H\over 3}a_0^2\phi_0^2\sqrt z\right]}.\ee

\subsubsection{ First order approximation}

Now for $n=-1$, equation (\ref{hbar1}) can be expressed as,
\be -\frac{\sqrt z}{36\beta x}\left(i S_{0,xx} - 2S_{0,x}S_{1,x} - \frac{i}{x}S_{0,x}\right) -\frac{1}{2 x z^{\frac{3}{2}}}\left(i S_{0,\phi\phi} - 2S_{0,\phi}S_{1,\phi}\right) + S_{1,z} - \frac{x^2}{48 \mathrm{H}^2 z^3}\phi S_{1,\phi} = 0. \ee
\noindent
Using the expression for $S_0$ in (\ref{S0}), we can write $S_{1,z}$ from the above equation as

\be S_{1,z} = \frac{i\left[A_0 z^2+ A_1 z + A_2\right]}{A_1 z^3+ A_3 z^2+{4\over 3}A_2 z}, \ee

where, $A_0 =\frac{(144\beta\mathrm{H}^2-12\alpha)}{a_0^2\phi_0^2}$,~~ $A_1 = \frac{\alpha}{96\beta\mathrm{H}^2}-{7\over 24}$, $A_2=\frac{a_0^2\phi_0^2}{1152\beta\mathrm{H}^2}$ and $A_3={4\over 3} - \frac{\alpha}{24\beta\mathrm{H}}$. On integration the form of $S_1$ in principle may be found as,

\be S_1 = i f(z).\ee
Hence the wavefunction up to first-order approximation is expressed as,
\be \psi = \psi_{01} e^{\frac{i}{\hbar}\left[-\frac{\alpha}{4}\mathrm{H} z^{\frac{3}{2}} + 48\beta\mathrm{H}^3 z^{\frac{3}{2}} -{\mathrm H\over 3}a_0^2\phi_0^2\sqrt z\right]},\ee
where,
\be \psi_{01} = \Psi_0 e^{f(z)}.\ee
Thus, first-order approximation only modifies the prefactor, keeping the oscillatory behavior of the wave function intact. The oscillatory behaviour of the wavefunction indicates that the region is classically allowed and the wavefunction is strongly peaked about a set of exponential solutions to the classical field equations. This establishes the correspondence between the quantum equation and the classical equations yet again.\\

\section{Summary}

In the present manuscript, we have attempted to resolve an age old debate regarding fixing of the end-point data for higher order theories of gravity. As already mentioned, canonical formulation of higher order theories require additional degree of freedom and hence additional end-point data.  While, some people insist on fixing the velocity at the end points \cite{rd2, rd3}, others believe that acceleration should be kept fixed, in addition to the coordinate \cite{10DH}. This issue has been discussed earlier in connection with generalized Pais-Uhlenbeck action \cite{ourpu}, where the authors in some details explored the advantage of fixing acceleration at the end-points, rather than the velocity. In the context of gravity, the question is whether $\delta h_{ij} = 0 = \delta K_{ij}$ or $\delta h_{ij} = 0 = \delta R$ at the boundary? In the introduction, although we have discussed several reasons to prefer the latter choice over the former, the example cited here, for minimally coupled higher-order gravity theory \eqref{AA} suggests none to prefer, since the two Hamiltonian (\ref{HO}/ \ref{hd11}/ \ref{hh}) and \eqref{hhl2} are canonically related. The situation ran worse in the case of non-minimal coupling \eqref{Aa1}. The two Hamiltonian (\ref{ho1}/ \ref{Hl2}/ \ref{Hl3}) and \eqref{3} are different, but not related under canonical transformation, while both are well posed. Fields associated with a particular system can not have two different phase-space structures. Thus, one of the techniques surely falls short. However, apparently, there is no way to prefer one over the other. This leads to degeneracy in Hamiltonian, which we shall discuss in a forthcoming article. Nevertheless, in connection with classical higher order (Pais-Uhlenbeck) oscillator \cite{ourpu}, we have demonstrated the advantages of fixing the acceleration at the end points, which is equivalent to fixing $\delta R = 0$ at the boundary in the case of higher-order theory of gravity. This is one of the reasons for modifying Horowitz' technique in the theory of gravity. Further, one can note that although in the case of minimal coupling \eqref{AA}, the two Hamiltonian are related under canonical transformation, still there is an additional important outcome of the Modified Horowitz' technique.  The Hamiltonian \eqref{hhl2} is associated with an effective potential in the form $V_e = {V(\phi)z\over x} + {3\alpha x\over 2 z}$, where, $z^{3\over 2} = a^3 = \sigma$ is the internal time parameter. If one now finds the extremum of the effective potential ${\partial V_e\over \partial x} = 0$, the result is $a = a_0 e^{\sqrt{ V(\phi)\over 6\alpha}t}$, using the relation $x = \dot z$. This is interesting, since it reveals that inflation is a generic feature of an action associated with higher order curvature invariant terms. This result would have remained obscure, unless Modified Horowitz' formalism were accounted for, since, on the contrary, the effective potential appearing form rest (\ref{HO}/ \ref{hd11}/ \ref{hh}) is $V_e = {2z\over 3 x} (V(\phi) + {\alpha^2\over 4\beta})$, which only yields $V(\phi) = -{\alpha^2\over 4\beta} = -V_0$, which is a constant. These are the essential causes to prefer Modification of Horowitz' formalism. The technique has been tested in anisotropic models as well \cite{11A31}. It is therefore possible to extend the technique beyond mini-superspace, provided the part of the boundary term ($\Sigma_{R_1}^2$) is handled appropriately in the whole super-space as well. This is of-course a future direction of research. \\

We would also like to draw the attention towards some recent developments in connection with Hamiltonian formulation of higher order theory of gravity in the form of conformal gravity \cite{rd1, rd2, rd3}. We briefly discuss the underlying difference with the present work. Deruelle et al \cite{rd1} studied Hamiltonian formulation of $f$(Riemann) theory of gravity which includes Weyl gravity. For the purpose, instead of the extrinsic curvature tensor $K_{ij}$, $\Omega_{ij}$ have been chosen as basic variables, which are essentially the components of Ricci tensor. The authors fixed the induced 3-metric $h_{ij}$ and the auxiliary variable $\Psi_{ij} = {1\over 2} {\partial f\over \Omega_{ij}}$ at the boundary, and at the end, linked their formalism with Ostrogradski's approach. First of all, although the action \eqref{AA} is a subclass of $f$(Riemann) theory of gravity, Gauss-Bonnet-Dilatonic coupled action \eqref{Aa1}, which is our main concern, falls outside its purview. Next, the authors admitted that the auxiliary variable in some cases, e.g. in the case of Lovelock action, becomes function of the extrinsic curvature tensor $K_{ij}$ and so the supplementary surface term considered by them, in general cannot be used to remove the second derivatives from the action. The same is true for Gauss-Bonnet-Dilatonic coupled case also. Finally, it is clear that the authors fix something else at the boundary which is proportional to $K_{ij}$ for Gauss-Bonnet-Dilatonic action and not the Ricci scalar. Equivalence with Ostrogradski's technique for standard action is obvious, and that is what we have explicitly demonstrated in section 2. Recently, $\mathrm{Kluso\check{n}}$ et al \cite{rd2} considered action with quadratic curvature terms, including the case of conformally invariant Weyl gravity and discussed the issue of boundary term in some detail. For making and breaking of conformal symmetry, Kiefer and Nikolic \cite{rd3} studied Weyl squared action, Weyl-Einstein gravity, and Weyl-Einstein gravity with nonminimally coupled scalar field, choosing unimodular conformal variables as the canonical ones. Again, Gauss-Bonnet-Dilatonic coupled action does not fall within the purview of the actions considered by these authors. Further, both have fixed extrinsic curvature at the end points, so boundary terms vanish. Finally, if the Hamiltonian so obtained can produce viable quantum theory has not been addressed in these works. \\

\end{document}